\documentclass[11pt,singlespace]{article}
\usepackage{putex}
\pdfoutput=1
\usepackage{graphicx}

\usepackage{hyperref}
\usepackage{subfigure}
\usepackage{graphicx}
\usepackage{enumerate}
\usepackage{cite}
\usepackage{tensor}
\usepackage{slashed}
\usepackage{placeins}
\usepackage{bigints}

\numberwithin{equation}{section}

\newcommand {\be} {\begin {equation}}
\newcommand {\ee} {\end {equation}}

\newcommand {\bes} {\begin {equation*}}
\newcommand {\ees} {\end {equation*}}

\newcommand{\beq}{\begin{equation}}
\newcommand{\eeq}{\end{equation}}

\newcommand{\vmoll}{ v_\text{M\o l}}

\newcommand{\erf}{\text{erf}}

\begin{document}

\institution{PU}{Department of Physics, Princeton University, Princeton, NJ 08544}

\title{
Lectures on Dark Matter Physics\\
}

\authors{Mariangela Lisanti 
}

\abstract{
Rotation curve measurements provided the first strong indication that a significant fraction of matter in the Universe is non-baryonic.  Since then, a tremendous amount of progress has been made on both the theoretical and experimental fronts in the search for this missing matter, which we now know constitutes nearly 85\% of the Universe's matter density.  These series of lectures, first given at the TASI 2015 summer school, provide an introduction to the basics of dark matter physics.  They are geared for the advanced undergraduate or graduate student interested in pursuing research in high-energy physics.   The primary goal is to build an understanding of how observations constrain the assumptions that can be made about the astro- and particle physics properties of dark matter.  The lectures begin by delineating the basic assumptions that can be inferred about dark matter from rotation curves.  A detailed discussion of thermal dark matter follows, motivating Weakly Interacting Massive Particles, as well as lighter-mass alternatives.  As an application of these concepts, the phenomenology of direct and indirect detection experiments is discussed in detail.}

\maketitle

Identifying the nature of dark matter (DM) remains one of the primary open questions in physics.  Measurements by Planck and WMAP demonstrate that nearly 85\% of the Universe's matter density is dark~\cite{Ade:2015xua}.  The Standard Model of particle physics alone cannot explain the nature of this DM, suggesting that the model must be extended.  All evidence in favor of particle DM thus far comes from observations of its gravitational effects on baryonic matter.  While we have amassed important clues from these results, many open questions remain:  What is the DM mass?  What is the strength of its interactions with visible matter?  How is it distributed throughout the Galaxy?  Fortunately, we are in the midst of a data-driven era in astroparticle physics that holds great promise towards addressing these questions.  A wide variety of experiments are currently reaching unprecedented sensitivity in their search for DM interactions in the lab and sky, and the field continues to evolve as new data forces re-evaluation of theory models.  

These lectures provide an introduction to DM physics for advanced  undergraduates and graduate students.  The primary goal is to help students build intuition for how to address the open questions in the field, while emphasizing the important interplay between theory and experiment.  We will begin by motivating a set of robust starting assumptions for the particle and astrophysical properties of DM and then show how these assumptions affect predictions for two classes of experiments: direct and indirect detection.  Exercises are interspersed throughout the text and provide an opportunity for the interested reader to reflect more deeply on the material.

\section{Astrophysical Distribution}

We begin our discussion by focusing on the astrophysical properties of DM.  This is a natural starting point because the strongest evidence for DM comes from its gravitational interactions with visible matter in the Milky Way.  Therefore, we will start with rotation curves, which provided the first robust clue for DM, and see just how much we can infer about the non-baryonic component of the Galaxy utilizing these observations.  Amazingly, this one piece of evidence is sufficient to infer the density and velocity distribution of DM in the Milky Way, and to posit the allowed mass range for the new matter particles.

Because the DM is essentially invisible to us, we must rely on visible objects that can act as tracers for it.  An adequate tracer must be collisionless, so that its distribution is determined primarily by its gravitational interactions, as should be the case for DM.  What are our options?  Well, the Milky Way contains approximately $\sim$$10^{11}$ stars with total mass $\sim$$5\times 10^{10}$ M$_\odot$~\cite{Binney}, where $M_\odot = 2.99\times10^{30}$~kg is a unit of solar mass.   The vast majority of these stars are located within the Galactic disk, which has a radius of $\sim$10~kpc and height of $\sim$0.5~kpc.  Our Sun, in particular, is located far out along one of the spiral arms, about 8.5~kpc from the Galactic Center where there is a black hole of mass $\sim$$4\times 10^6$~M$_\odot$.  The interstellar medium, composed primarily of atomic and molecular Hydrogen, is concentrated along the Galactic disk, and makes up roughly 10\% of the total stellar mass.    

Among these options, it turns out that stars serve as the best DM tracers and time-and-again have provided important clues about its distribution.  The reason for this is that stars in the disk are essentially collisionless.  The time between collisions is $\sim$$10^{21}$ years---far longer than the age of the Universe!

\vspace{0.2in}
\vbox{
\hrule
\vspace{0.1in}
\noindent \textsc{Exercise:} Show that the stars in the Galactic disk are collisionless.  You may assume that the stars have a radius similar to that of the Sun ($R_\odot = 2\times 10^{-8}$~pc) and a random velocity of $\sim$50~km/s.  
\vspace{0.1in} 
\hrule
}
\vspace{0.1in}

\subsection{Rotation Curves} 
 
\begin{figure}[tb] 
   \centering
   \includegraphics[width=0.9\textwidth]{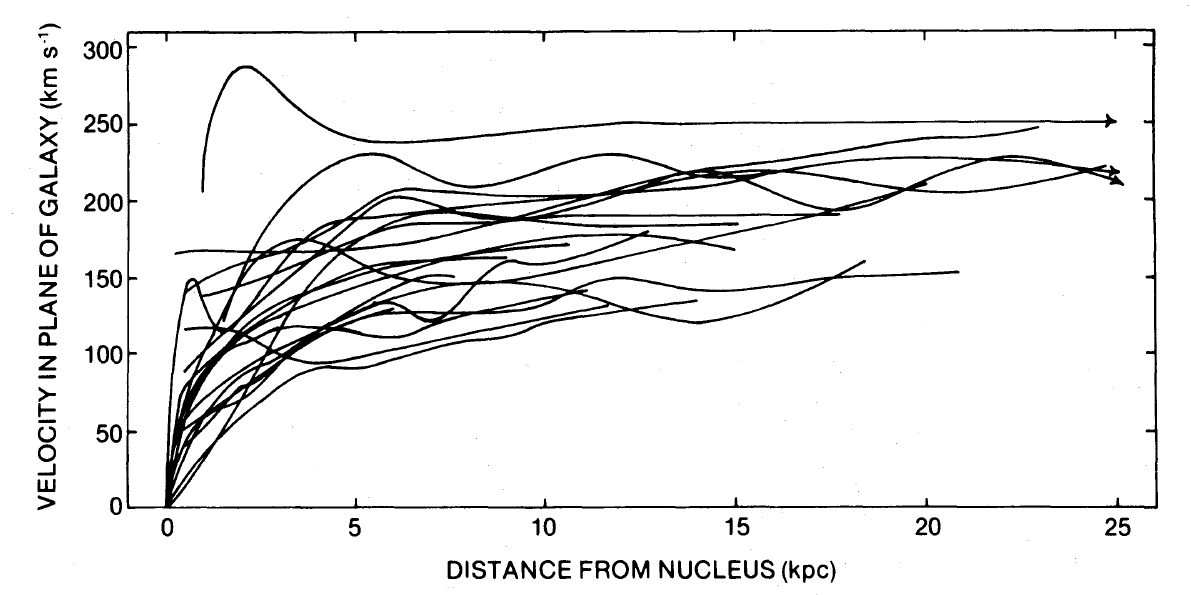} 
   \caption{Rotation curves of spiral galaxies as measured in the original Rubin \emph{et al.} paper~\cite{Rubin:1980zd}.  Most galaxies show a flattening of the circular velocity at large radial distances. }
   \label{fig: M31rotcurve}
\end{figure}

One of the strongest pieces of evidence for DM comes from studying the rotational velocity of stars.  The fact that stars rarely collide means that their motion is dictated by their gravitational interactions.  From standard Newtonian gravity, we know that the stars' circular velocity, $v_c$, is 
\begin{equation}
v_c(r) = \sqrt{\frac{GM}{r}} \, , \nonumber
\end{equation}
where $M$ is the enclosed mass, $r$ is the radial distance,  and $G$ is the gravitational constant.  For distances that extend beyond the Galactic disk ($r\gtrsim R_\text{disk}$), Gauss' Law tells us that $M$ should remain constant assuming all the mass is concentrated in the disk, and $v_c \propto r^{-1/2}$.  Instead, observations find that the circular velocity curve flattens out at these distances, implying that $M(r) \propto r$.  This suggests that there is an additional `dark' component of matter beyond the visible matter in the disk.\footnote{Another  interpretation for the flattening of the rotation curve is the possibility that  Newton's laws of gravity are altered at large distances~\cite{Milgrom}.  MOdified Newtonian Dynamics (MOND) is a class of  phenomenological models that seek to address this point.  While MOND is most successful at explaining galaxy-scale effects, it has not been absorbed into a fully cosmological picture to date.}  Evidence for flat rotation curves began to build in the 1970s (\emph{e.g.},~\cite{RubinFord,Whitehurst}), leading to several ground-breaking papers in the early 1980s~\cite{Rubin:1980zd,Bosma:1981zz}.  Figure~\ref{fig: M31rotcurve} shows the 21 Sc rotation curves measured by Rubin \emph{et al.} in~\cite{Rubin:1980zd}, which illustrate the approximate flattening of the circular velocity at large radial distances.  Since then, further evidence has continued to strengthen these conclusions---see \emph{e.g.},~\cite{Persic:1995ru}.

From rotation curves, we infer that the DM mass density distribution is 
\be
\rho(r) \propto \frac{M(r)}{r^3} \sim \frac{1}{r^2} \, .  \nonumber
\ee
Note the implicit assumption being made here: namely, that the DM is distributed in a spherically symmetric halo about the center of the Galaxy, in contrast to the baryons which are concentrated in the disk.  Because baryons can interact strongly amongst themselves, they have a means of dissipating energy and can thus collapse into a disk.  DM, in contrast, is not dissipative\footnote{Theories with several DM components may include a small fraction of self-interacting particles that collapse to form a disk~\cite{Fan:2013yva}.} and thus forms spherical `halos.'

Let us get an order-of-magnitude estimate for the mass and size scale of the Milky Way's DM halo.  Stellar kinematics constrain the total mass of the halo to be $M_\text{halo}\sim$$10^{12}$ M$_\odot$ and the local DM density to be $\rho_0 \sim 0.3$~GeV/cm$^3$.  For a review of the current status of these measurements and a detailed discussion of the (large) uncertainties, see~\cite{Read:2014qva}.  Therefore, the radius of the halo, $R_\text{halo}$, is approximately
\be
M_\text{halo} \sim 4 \pi \int_0^{R_\text{halo}} dr \, r^2 \rho(r)  \longrightarrow R_\text{halo} \sim 100~\text{ kpc} \, , \noindent \nonumber
\ee
taking the mass density motivated by rotation curves.  It is worth stressing that this is only a very rough estimate for $R_\text{halo}$.  It assumes a spherically symmetric density distribution that is almost certainly too simplistic.  Also, it is not correct to think of a halo as having a finite end and establishing an adequate reference for comparing the sizes of different halos is a subtle point that we will not address here---see~\cite{White2001}.  All that being said, the estimate above indicates that the DM halo extends out roughly an order of magnitude beyond the baryonic disk!

The average velocity of the DM in the halo can be obtained using the virial theorem:
\be
\langle v \rangle \sim \sqrt{\frac{G M_\text{halo}}{R_\text{halo}}} \sim 200~\text{ km/s} \, . \nonumber
\ee
Importantly, notice that the DM is non-relativistic---this will end up playing an important role in predicting observational signatures.  We can further refine our estimates of the expected DM velocity by taking advantage of the fact that the DM density and velocity are related through the gravitational potential.  As a result, once the density distribution is set, there is an associated velocity distribution that makes the theory self-consistent.  In particular, an isotropic halo in steady state with an inverse-square density distribution has a Maxwellian velocity distribution.  Deriving this explicitly requires knowledge of the Boltzmann equation and Jeans Theorem, so let us take a moment to review these important concepts.  

\subsection{The Boltzmann Equation \& Jeans Theorem}

The Boltzmann equation describes the evolution of the phase-space density $f( \textbf{x}, \textbf{v})$ of a DM particle in the halo.  This gives the probability  $f( \textbf{x}, \textbf{v})\,d^3\textbf{x} \, d^3\textbf{v}$ of finding the particle in some volume $d^3\textbf{x} \, d^3\textbf{v}$.  Conservation of probability dictates that
\be
\int f( \textbf{x}, \textbf{v}) \, d^3\textbf{x} \, d^3\textbf{v} = 1 \, .\nonumber 
\ee
The Boltzmann equation states that 
\be
\textbf{L}[f] = \textbf{C}[f] \, ,
\label{eq: Boltz}
\ee
where \textbf{L} and \textbf{C} are the Liouville and collision operators, respectively.  The most general form for the Liouville operator is
\be
\textbf{L}[f] = p^\alpha \, \frac{\partial f}{\partial x^\alpha} - \Gamma_{\beta \gamma}^{\alpha} \, p^\beta \, p^\gamma \, \frac{\partial f}{\partial p^\alpha} \, ,
\label{eq: L}
\ee
  where $\Gamma^\alpha_{\beta \gamma}$ is the affine connection.  In the non-relativistic limit, (\ref{eq: L}) simplifies to 
\be
\textbf{L}_\text{nr}[f] = \frac{\partial f}{\partial t} + \dot{\textbf{x}} \, \frac{\partial f}{\partial \textbf{x}} +  \dot{\textbf{v}} \, \frac{\partial f}{\partial \textbf{v}} \, . \nonumber
\ee
The collision operator $\textbf{C}[f]$ includes interactions between DM and other particles (including itself) that may alter the phase-space density.  This operator can take a complicated form, depending on the allowed interactions of the particles.  To model the phase-space distribution of DM in the Milky Way today, however, we can work with a simple form of the Boltzmann equation because the DM in the halo is non-relativistic and collisionless, so 
\be
 \frac{\partial f}{\partial t} + \dot{\textbf{x}} \, \frac{\partial f}{\partial \textbf{x}} +  \dot{\textbf{v}} \, \frac{\partial f}{\partial \textbf{v}} = 0 \, .
 \label{eq: collBoltz}
\ee

The collisionless Boltzmann equation has only a restricted set of solutions for $f(\mathbf{x},\mathbf{v})$.  Jeans Theorem states that a steady-state solution to (\ref{eq: collBoltz}) can only be a function of the phase-space coordinates through the integrals of motion $I(\textbf{x}, \textbf{v})$, which satisfy
\be
\frac{d}{dt} I (\textbf{x}(t) , \textbf{v}(t)) = 0 \, . \nonumber 
\ee
For a detailed discussion of Jeans Theorem, see~\cite{Binney}.  The Hamiltonian is one example of an integral of motion.  In this case, Jeans Theorem tells us that the phase-space distribution is solely a function of energy $\mathcal{E}$ for a halo in steady state:
\be
f(\mathbf{x}, \mathbf{v}) = f(\mathcal{E}) \, \text{ where} \quad \mathcal{E} = \Psi - \frac{1}{2} v^2 \, \nonumber
\ee
and $\Psi$ is the gravitational potential.  Other integrals of motion include the angular momentum variables $\mathbf{L}$ or $L_z$.  Distribution functions such as $f(\mathcal{E}, \mathbf{L})$ or $f(\mathcal{E}, L_z)$ are all allowed solutions to (\ref{eq: collBoltz}), leading to different velocity distributions in each case.

\vspace{0.2in}
\vbox{
\hrule
\vspace{0.1in}
\noindent \textsc{Exercise:} Show that the mean DM velocity $\langle \mathbf{v} \rangle$ vanishes and that the velocity-dispersion tensor $\langle v_i v_j \rangle$ is isotropic if the phase-space distribution is only a function of energy.
\vspace{0.1in} 
\hrule
}

For an isotropic halo in steady-state with velocity distribution $f(\mathcal{E}) \propto e^\mathcal{E}$, the associated density distribution is
\be
\rho \propto \int_0^\infty dv \, v^2 \, f(v) = \int_0^\infty dv \, v^2 \, \exp \left(\frac{\Psi - v^2/2}{\sigma^2}\right) \propto e^{\Psi/\sigma^2} \, ,\nonumber
\ee
where $\sigma$ is the velocity dispersion.  The above expression highlights very clearly the interrelation between the density and velocity distributions, through their dependence on $\Psi$.  Using Poisson's equation and the fact that $\Psi \propto \ln \rho$, we solve for the radial dependence of the density distribution:
\be
\nabla^2 \Psi = -4 \pi G \rho \longrightarrow \rho(r) = \frac{\sigma^2}{2\pi G r^2} \, .\nonumber
\ee
Therefore, the phase-space distribution for a spherical isotropic halo in steady state is well-modeled by 
\be
\rho(r) \propto 1/r^2  \quad \text{ and } \quad f(v) \propto e^{-v^2/\sigma^2} \, .  \nonumber
\ee
This is precisely what is expected for a self-gravitating isothermal gas sphere!  The fact that rotation curves motivate an inverse-square fall-off for the density distribution appears to provide support for this scenario.  However, while it provides intuition, this simple picture must be augmented because such a density profile predicts an infinitely massive halo.  This in turn suggests that at distances beyond current measurements, rotation curves must no longer be approximately flat.

\subsection{Input from Numerical Simulations}

Our estimates above relied on the fact that the Galaxy is in a steady state, which means that the virial theorem or Jeans Theorem must apply.  This is a reasonably good approximation for the Milky Way.  Observational evidence suggests that the Milky Way's last major merger with another galaxy occurred $\sim$10~Gyr ago (compared to our Galaxy's 13~Gyr age)---see \emph{e.g.}, \cite{Gilmore:2002jv, Abadi:2002tt}.  A major galaxy merger is one that significantly perturbs the Galactic disk, distorting the spiral structure.  However, minor mergers are  continuing to this day, as evidenced most spectacularly by the Sagittarius dwarf galaxy~\cite{Ibata:1994fv}.  As this dwarf galaxy has been falling into the Milky Way, it has left a tail of tidal debris in its wake, which is observed as a stream of stars in the sky that roughly traces the expected orbit of the infalling galaxy~\cite{Ivezic:2000ua,Yanny:2000ty, Ibata:2000pu,Ibata:2001iw,Johnston:1995vd}.  This suggests that the steady-state assumption is only approximately correct and that the merger history of the Galaxy does affect the phase-space distribution of DM.    

To address this point properly requires leaving the realm of analytic (or semi-analytic) calculations and turning to numerical simulations that properly model the hierarchical merging of individual DM halos.  These simulations follow structure formation, from the initial DM density perturbations to the largest halos today.  They are referred to as N-body simulations because they keep track of the many-body gravitational interactions between DM halos as they merge together to form ever larger structures.  To date, the highest resolution simulations that model Milky-Way--like halos (\emph{e.g.}, \texttt{Via Lactea II}~\cite{Diemand:2008in} and \texttt{Aquarius}~\cite{Springel:2008cc}) only contain DM.  However, the inclusion of baryons is likely to have important consequences.  Simulations that account for the gas physics are computationally more expensive but can result in distinct changes to the DM density and velocity distributions, especially in baryon-rich regions.  For example, adiabatic contraction leads to condensation of gas towards the center of the halo, which can pull the DM with it, enhancing the halo's central density~\cite{Blumenthal:1985qy,Gnedin:2004cx}.  In contrast, complex feedback mechanisms from \emph{e.g.}, energy injection from Active Galactic Nuclei and supernova outflows can eject DM away from the center of the halo--- see \emph{e.g.},~\cite{Mashchenko:2007jp, Governato:2009bg}.  As the numerical resolution of full hydrodynamic simulations continues to improve, we will gain a better understanding of which of these two mechanisms wins out in our own Galaxy.

\begin{figure}[tb] 
   \begin{center}$
	\begin{array}{cc}
	\scalebox{0.28}{\includegraphics{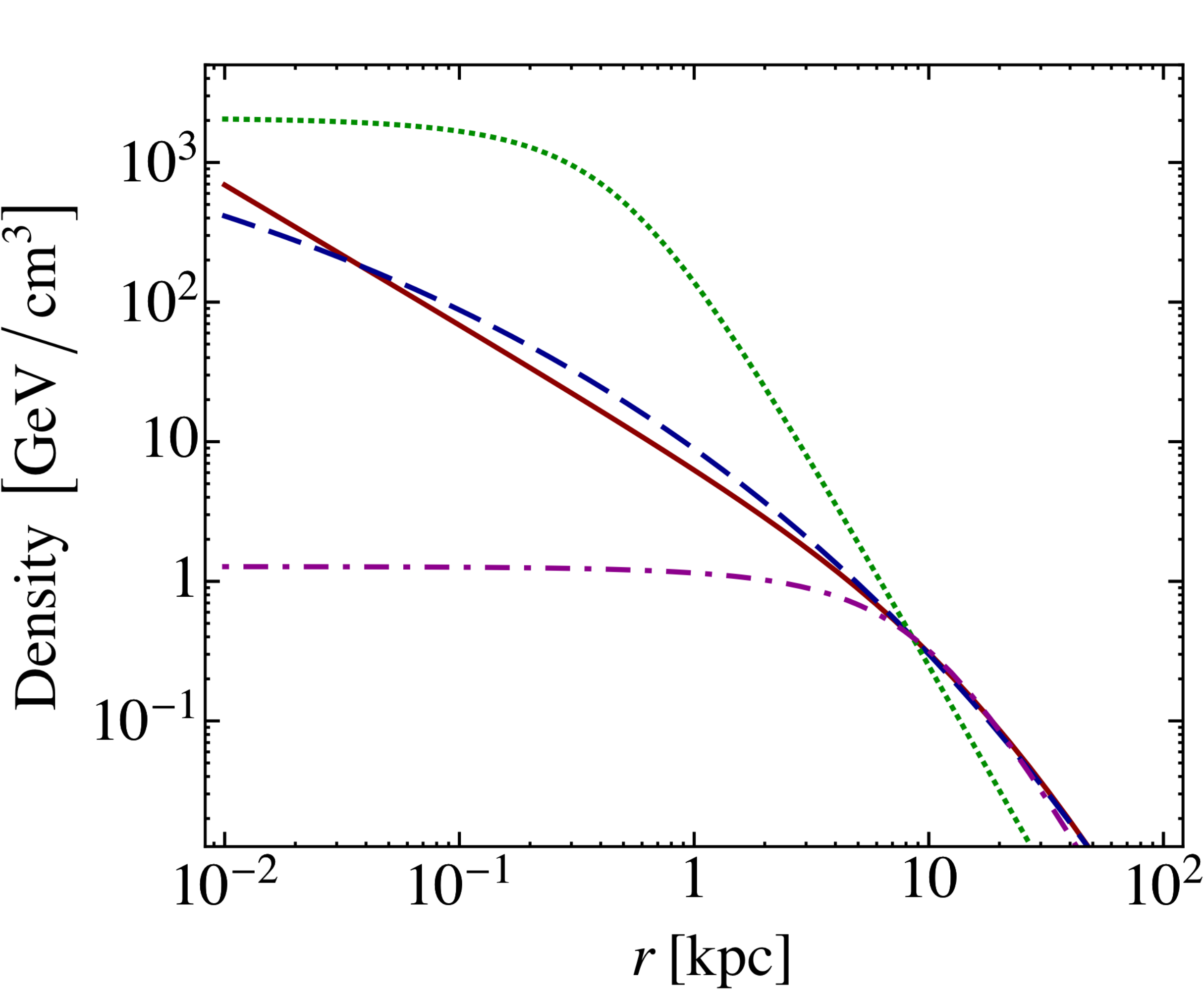}} & \scalebox{0.8}{\includegraphics{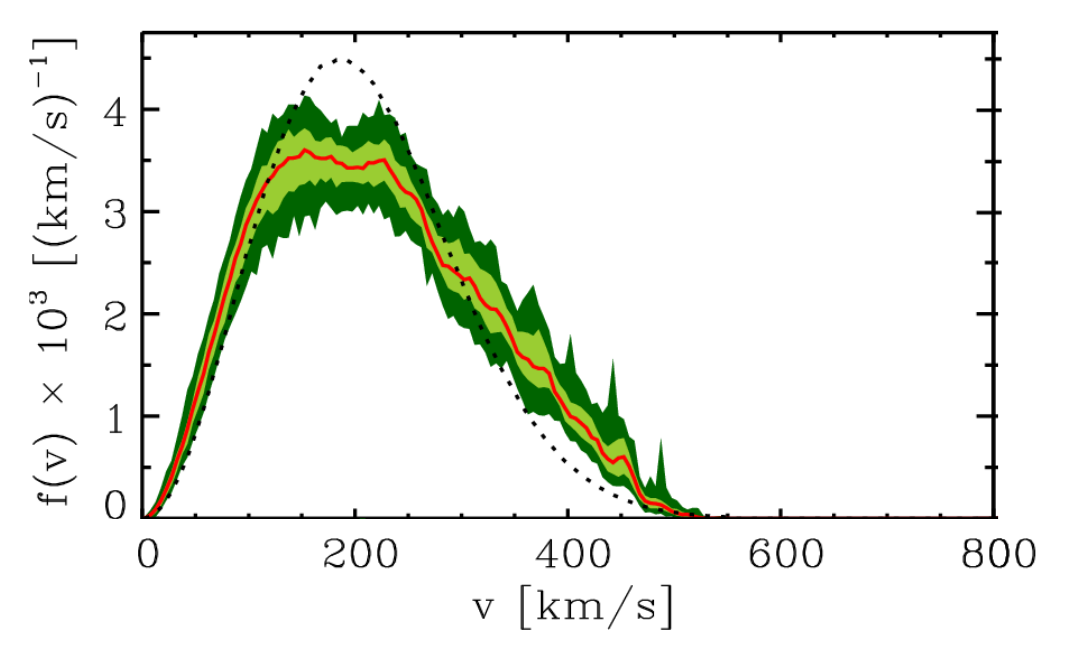}} 
	\end{array}$
	\end{center}
   \caption{\emph{(left)} A comparison of the NFW (solid red), Einasto (dashed blue), and Burkert with $r_s = 0.5$ (dotted green) and 10~kpc (dot-dashed purple) profiles.  Figure from~\cite{Cohen:2013ama}.  \emph{(right)}  The expected velocity distribution from the \texttt{Via Lactea} simulation (solid red), with the 68\% scatter and the minimum/maximum values shown by the light and dark green shaded regions, respectively.  For comparison, the best-fit Maxwell-Boltzmann distribution is shown in dotted black.  Figure from~\cite{Kuhlen:2009vh}.}
   \label{fig: profiles}
\end{figure}

What have we learned about the DM's density and velocity distributions from simulations so far?  First, the density distribution appears to be approximately universal and well-modeled by the Navarro-Frenk-White (NFW) profile~\cite{Navarro:1995iw}:
\be
\rho_\text{NFW}(r) = \frac{\rho_0}{r/r_s (1+r/r_s)^2} \, ,  \nonumber
\ee
where $r_s=20$~kpc is the scale radius.  Note that the radial dependence is different than that for the isothermal profile, highlighting that our simple estimates from above only approximately reconstruct the more complete picture.  The Einasto profile, which also appears frequently in the literature, takes the form: 
\be
\rho_\text{Ein}(r)  = \rho_0 \exp \left[ -\frac{2}{\gamma} \left( \left(\frac{r}{r_s}\right)^\gamma-1\right) \right] \, , \nonumber
\ee
with $r_s = 20$~kpc and $\gamma = 0.17$~\cite{Einasto}.  While both NFW and Einasto are preferred by DM-only simulations, it is possible that the story changes in full hydrodynamic simulations.  It may be possible that the inner profile is more cored (\emph{e.g.}, has a flatter slope) than the NFW or Einasto profiles, which are described as `cuspy' because of their steeper inner slopes.  The Burkert profile~\cite{Burkert:1995yz} is one such example:
\be
\rho_\text{Burk}(r) = \frac{\rho_0}{(1+r/r_s) (1+(r/r_s)^2)} \, , \nonumber
\ee
where $r_s$ is the core radius.  A comparison of the NFW, Einasto, and Burkert profiles is shown in the left panel of Fig.~\ref{fig: profiles}.  Observational evidence from dwarf galaxies (small galaxies with few stars) may suggest cored profiles (see \emph{e.g.},~\cite{Walker:2011zu}).  While baryonic feedback mechanisms may suffice in explaining such cored profiles~\cite{Brooks:2014qya}, they may also be due to another cause all-together---non-trivial DM self-interactions~\cite{Spergel:1999mh}.  

The fact that the density distribution recovered from N-body simulations differs from isothermal tells us with certainty that the Maxwell-Boltzmann distribution is not the correct velocity distribution.  Remember that the density and velocity distributions must be self-consistent, as they are related to each other through the gravitational potential, and a Maxwellian velocity distribution requires $\rho \propto r^{-2}$.  The right panel of Fig.~\ref{fig: profiles} compares the velocity distribution obtained from the \texttt{Via Lactea} N-body simulation with the Maxwell-Boltzmann distribution.  Notice that the \texttt{Via Lactea} distribution has more high-speed particles relative to the Maxwellian case.  Debate continues as to how this conclusion changes in full hydrodynamic simulations~\cite{Bozorgnia:2016ogo,Kelso:2016qqj,Sloane:2016kyi}.  However, the important point to make is that the \emph{tail} of the velocity distribution is most sensitive to the merging history of the halo.  When a subhalo falls into the Galaxy, it is tidally disrupted and leaves behind remnants that are out of equilibrium.  The DM particles in these remnants are likely to have higher speeds, on average, than the rest of the halo and will contribute to the high-velocity tail of the velocity distribution.  Therefore, the shape of the high-velocity end of the distribution depends on the size and time of minor mergers in our own Galaxy.  

Despite the caveats listed here, the distribution that is used most often in the literature is the truncated Maxwellian, otherwise known as the Standard Halo Model:
\be
   f(\textbf{v}) = \left\{
     \begin{array}{lr}
       \frac{1}{N_\text{esc}} \left( \frac{3}{2\pi \sigma_v^2} \right)^{3/2} e^{-3 \mathbf{v}^2/2\sigma_v^2}  & : |\mathbf{v}| < v_\text{esc} \\
       0 & : \text{otherwise} \\ 
            \end{array} \nonumber
   \right.
\ee
where $\sigma_v$ is the rms velocity dispersion, $v_0 = \sqrt{2/3} \sigma_v \approx 235$~km/s is the most probable speed~\cite{Kerr:1986hz,Reid:2009nj,McMillan:2009yr,Bovy:2009dr}, and $N_\text{esc} = \erf(z) - 2 \pi^{-1/2} z e^{-z^2}$, with $z\equiv v_\text{esc}/v_0$ and $v_\text{esc}$ the escape velocity.  

N-body simulations also find evidence for substructure in the DM phase-space distribution.  This includes localized features that arise from relatively recent minor mergers between the Milky Way and other galaxies.  When another DM subhalo falls into an orbit about the center of the Milky Way, tidal effects strip DM (and, possibly, stars) along its orbit.  This `debris' eventually virializes with the other particles in the Milky Way's halo.  However, at any given time, there is likely to be some fraction of this debris that has not come into equilibrium and which exhibits unique features that may affect observations.  
\begin{figure}[tb] 
   \centering
   \includegraphics[width=5in]{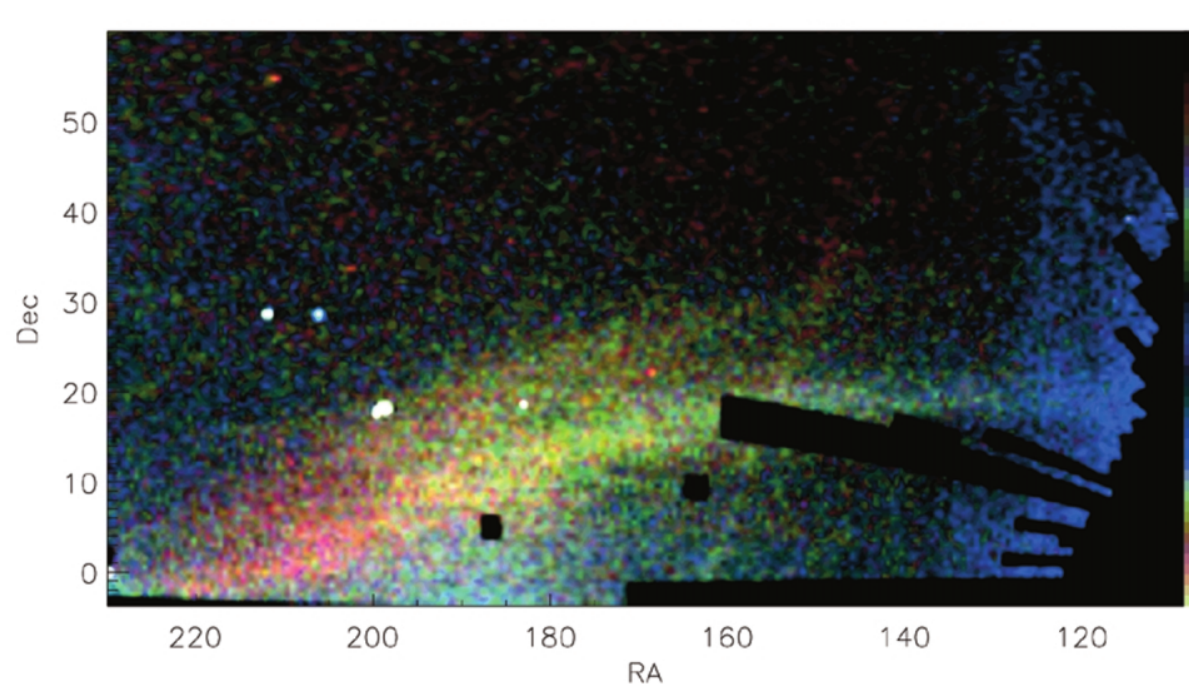} 
   \caption{The Field of Streams is a patch of sky imaged by SDSS that contains several overlapping stellar streams~\cite{Belokurov:2006ms}.  The color denotes distance of the stars, with blue the most nearby and red the farthest.  The yellow and red band shows two tails of the Sagittarius stream (note the bifurcation at RA$\sim$180$^\circ$).  The Monoceros stream is the vertical blue swath to the right of the image.  The Orphan stream is the thin vertical stripe near RA$\sim$160$^\circ$.     }
   \label{fig: FieldOfStreams}
\end{figure}
Examples of substructure include:
\begin{itemize}
\item \textbf{Clumps}:  Concentrated clumps of DM may be left behind by the merging process.  Each clump would result in a localized overdensity of DM. 
\item \textbf{Streams}: A tidal stream is an example of debris left behind along the orbits of infalling subhalos.  Figure~\ref{fig: FieldOfStreams} is a famous image from the Sloan Digital Sky Survey (SDSS) known as the `Field of Streams.'  The single patch of sky in this image contains several arms of the Sagittarius stream, as well as the Orphan and Monoceros stellar streams.  Evidence for stellar streams suggests that similar features might form in the DM distribution as well.  If this were the case, then the DM velocities in a given stream would be coherent, with
\be
f_\text{stream}(\mathbf{v}) = \delta^{(3)}\left(\mathbf{v} - \mathbf{v}_\text{stream} \right) \, .\nonumber
\ee
The right panel of Fig.~\ref{fig: profiles} shows localized spikes in the tail of the velocity distribution, which are associated with streams in \texttt{Via Lactea}. 
\item \textbf{Debris Flow}:  Imagine what happens as many subhalos orbit about the Milky Way, each dumping tidal debris along its path.  Debris flow is the sum total of the overlapping streams, shells, and plumes of debris from these mergers.  The aggregate sum of this partially virialized material is spatially homogenous, resembling a fluffy cloud.  Despite being spread out in position-space, its constituents retain structure in velocity-space.  In particular, they share a common speed, even though their velocities are not coherent in direction~\cite{Lisanti:2011as,Kuhlen:2012fz}.  Simulations suggest that debris flow may comprise a significant fraction of the high-speed particles in the Milky Way.  In the right panel of Fig.~\ref{fig: profiles}, debris flow accounts for the excess of particles on the tail of the \texttt{Via Lactea} distribution, relative to the Maxwellian expectation.   
\item \textbf{Dark Disk}:  N-body simulations suggest that a dark disk may form if subhalos merging with the Milky Way are dragged through and disrupted by the baryonic disk~\cite{Read:2009iv,Read:2008fh,Purcell:2009yp,Bruch:2008rx}.  The net result is a concentration of DM along the plane that rotates in the same direction as the Sun, except with a lag speed of $\sim$50~km/s.  A dark disk would enhance the local DM density and also provide an excess of slower-moving DM particles in the Solar neighborhood.  Simulations suggest that a dark disk can lead to a factor of 0.5--2 overdensity in the local density, however observations may constrain this further~\cite{Read:2014qva}.
\end{itemize}

\section{Particle Physics Properties}

The previous section motivated properties of the astrophysical distribution of DM from measurements of rotation curves.  As it turns out, it is possible to make some general, model-independent statements about the mass range by simply requiring that DM form halos.  The lower bound of allowed masses is set by the number of particles that can be confined within a given cell of phase space, which is set by the spin statistics of the particle.  For example, if the DM is an ultra-light scalar, then Bose statistics dictates that there is no limit to the number of particles that can be packed into the same point in phase space.  In this case, the occupation number of DM particles is so high that it can effectively be treated as a classical field and the stability of the halo is set by the uncertainty principle using $\Delta x \, \Delta p \sim  1$, where $\Delta p \sim m_\chi v$ and $\Delta x\sim 2 R_\text{halo}$.  The tightest bounds come from halos surrounding dwarf galaxies, from which we estimate that a scalar DM particle must have mass greater than
\begin{equation}
m_\text{scalar} \gtrsim 10^{-22}~\text{eV} \,. \nonumber
\end{equation}
Ultra-light scalar DM particles near the bottom of this bound are referred to as `fuzzy' dark matter~\cite{Hu:2000ke}.  

The argument changes for fermions due to Pauli exclusion~\cite{TremaineGunn,Madsen:1991mz,Zee}.  This means that
\be
M_\text{halo} = m_\text{ferm} V \, \int f(p) \, d^3p \, \lesssim \, m_\text{ferm} V \int d^3 p \sim m_\text{ferm} \, R_\text{halo}^3 \, \left(m_\text{ferm} v\right)^3 \, , \nonumber
\ee
where $V= \frac{4}{3} \pi R^3$ is the volume of a spherical halo of radius $R$ and $m_\text{ferm}$ is the mass of the fermionic particle.  The $\lesssim$ arises from the fact that each unit volume of phase space can have up to, but no more than, one fermionic particle, on average.  Substituting in for the virial velocity gives
\be
m_\text{ferm} \gtrsim \left( G^3 M_\text{halo} R_\text{halo}^3 \right)^{-1/8} \,\nonumber
\ee
and $m_\text{ferm} \gtrsim \mathcal{O}(10)$~eV.  Generalizations of this phase-space argument lead to even tighter constraints.  For example, the phase-space densities of dwarf galaxies suggest that~\cite{Horiuchi:2013noa}  
\be
m_\text{ferm} \gtrsim 0.7 \text{ keV} \,. \nonumber
\ee
As expected, the bound on fermionic DM is much more stringent than that for bosonic DM.  

To stress, these restrictions on the DM mass range are the most \emph{generic} statements that can be made.\footnote{The upper mass limit for DM comes from searches for MACHOs, MAssive Compact Halo Objects.  Such objects cause lensing events when they pass in front of bright stars and the lack of such detections excludes MACHOs with masses between roughly $10^{57-67}$~eV~\cite{Griest:2013esa, Tisserand:2006zx,EROS}.}  However, folding in assumptions about the evolution of the DM density in the early Universe can motivate more specific mass scales.  
\begin{figure}[tb] 
   \centering
   \includegraphics[width=5in]{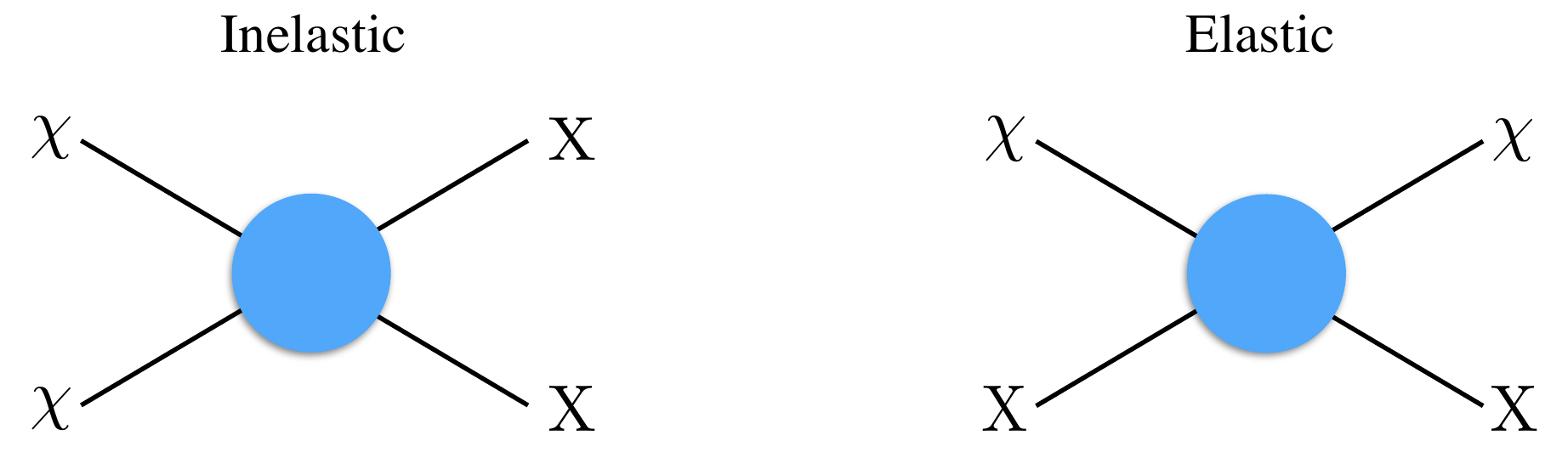} 
   \caption{An illustration of the inelastic $\chi \chi \rightarrow XX$ (left) and elastic $\chi X \rightarrow \chi X$ (right) scattering processes that dictate chemical and kinetic equilibrium.  Note that time points from left to right. }
   \label{fig: equil}
\end{figure}
Let us now focus on the case where a DM particle is in thermal equilibrium in the early Universe via its interactions with Standard Model particles.

\subsection{Thermal Dark Matter}

Figure~\ref{fig: equil} shows two possible $2\rightarrow2$ interaction diagrams that are allowed with $\chi$ the DM particle and $X$ a Standard Model particle, which is essentially massless and in equilibrium with the photon bath.  When the interaction $\chi \chi \leftrightarrow X X$ is in equilibrium, the DM particles are constantly being replenished.  As the Universe expands, though, it becomes increasingly harder for a DM particle to find a partner to annihilate with and the forward reaction shuts off.  At this point, the DM density remains frozen in time.  The `freeze-out' time occurs when the annihilation rate, $\Gamma_\text{inelastic}$, is on the order of the Hubble rate, $H$:
\be
\Gamma_\text{inelastic} = n_\chi \langle \sigma v \rangle \sim H \, ,  \nonumber
\ee
where $n_\chi$ is the DM number density and $\langle \sigma v \rangle$ is the velocity-averaged cross section.  Cold DM is non-relativistic at freeze-out, with $n_\chi \sim T^{3/2} e^{-m_\chi/T}$, with $T$ the temperature of the DM species; hot DM is relativistic at freeze-out, with $n_\chi \sim T^{3}$.  Warm DM falls somewhere in between these two cases.

\vspace{0.2in}
\vbox{
\hrule
\vspace{0.1in}
\noindent \textsc{Exercise:} The number density of a given particle is related to its phase-space density, $f(E,t)$, via
\be
n = g\int f(E,t) \, \frac{d^3 p}{(2 \pi)^3} \, , 
\label{eq: numdens}
\ee
where $g$ is the number of spin degrees of freedom of the particle.  Determine the scaling of $n$ with temperature in the non-relativistic and relativistic limits.
\vspace{0.1in}
\hrule
}
\vspace{0.2in} 
\begin{figure}[tb] 
   \centering
   \includegraphics[width=3.5in]{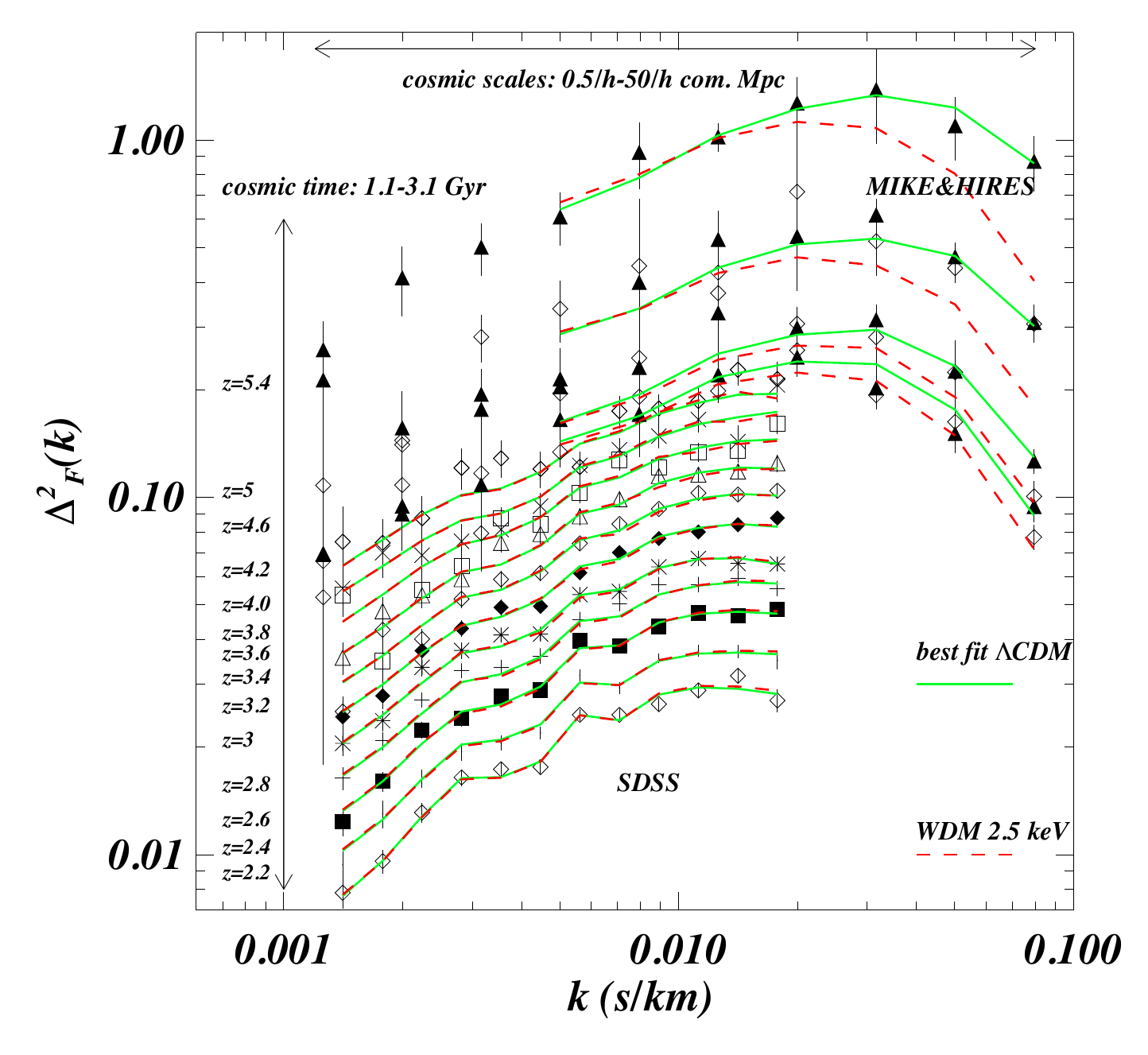} 
   \caption{Lyman-$\alpha$ flux power spectra for time slices that span 1.1--3.1~Gyr after the Big Bang.  The best-fit cold(warm) DM fits are shown in solid green(dashed red).  The warm DM curves do a poor job at reproducing the data at high redshift.  Figure from~\cite{Viel:2013apy}.}
   \label{fig: powerspectrum}
\end{figure}

After freeze-out, the DM is no longer in chemical equilibrium, but it remains in thermal equilibrium with the surrounding plasma via the elastic interaction shown in the right panel of Fig.~\ref{fig: equil}.  After a certain point, however, even this interaction decouples.  The elastic interaction rate,
\be
\Gamma_\text{elastic} = n_X \langle \sigma v \rangle \, ,  \nonumber
\ee
is proportional to $n_X$, which scales like $T^3$ as the $X$ are relativistic.  For cold DM (CDM), $\Gamma_\text{elastic}$ exceeds the Hubble rate only after the DM has fallen out of chemical equilibrium; at this point in time (referred to as `kinetic decoupling'), the DM is free streaming.  In contrast, kinetic decoupling happens earlier for hot DM.

The size of the horizon during kinetic decoupling sets a cutoff scale for the DM power spectrum.  Before decoupling, the DM fluid is coupled to the photon bath and perturbations are damped by friction between the two.  When the DM free streams after decoupling, it experiences collision-less damping because the particles move in random directions due to a non-zero average velocity.  Detailed modeling of these effects is somewhat involved (see the review~\cite{Bringmann:2006mu} and references therein), but the net effect is to strongly suppress the perturbation spectrum below some characteristic wavenumber (or above some free-streaming length).  The hotter the DM, the lower the cutoff because its free-streaming length is larger after decoupling.  Figure~\ref{fig: powerspectrum} shows the Lyman-$\alpha$ power spectra measured from 25 different high-redshift quasars in~\cite{Viel:2013apy}.  Notice that the best-fit warm DM curves poorly reproduce the power spectra at $z \gtrsim 5$.  The study excludes thermal DM candidates (that comprises 100\% of the DM density) for masses $m_\text{thermal} \gtrsim 3.3~\text{keV}$ at 2$\sigma$ confidence.  These results constrain warm DM candidates, which predict less structure on small scales than is actually observed.

Warm DM can potentially explain some inconsistencies between CDM simulations and observations on galactic scales.  The three most often-cited challenges are:~(1) The \emph{Missing Satellites Problem} where N-body simulations predict more satellite galaxies in orbit around the Milky Way than are actually observed.~(2) The \emph{Cusp/Core Controversy} where some data from dwarf galaxies point to a shallow central slope of the density profile, not recovered in DM-only N-body simulations.~(3) The \emph{Too Big to Fail Problem} in which we do not observe dwarf galaxies that are as large as the ones found in simulations.  Warm DM can help to resolve the first two of these challenges, in particular, because its longer free-streaming length (relative to CDM) washes out structure on these scales.  Debate continues as to whether these challenges can be resolved with full hydrodynamic N-body simulations that properly include the feedback from baryonic processes~\cite{Brooks:2014qya}.

\subsection{Freeze Out}

For the remainder of this section, we will delve more deeply into the CDM scenario, specifically.  To calculate the DM number density today, we follow the evolution of the inelastic scattering process with time using the Boltzmann equation (\ref{eq: Boltz}).  This application of the Boltzmann equation requires the covariant form of the Liouville operator, which can be written as 
\be
L[f] = E \frac{\partial f}{\partial t} - \frac{\dot{a}}{a} | \textbf{p} |^2 \frac{\partial f}{\partial E} \, .  \nonumber
\ee
Using (\ref{eq: numdens}), 
\be
g \int L[f] \, \frac{d^3 p}{(2 \pi)^3} = \frac{1}{a^3} \, \frac{d}{dt}\left(na^3\right)= \frac{dn}{dt} + 3 \, H \, n \, ,
\label{eq: Lcov}
\ee    
where $H=\dot{a}/a$ is the expansion rate of the Universe and $a$ is the scale factor.  When there are no number-changing DM interactions (that is, when $\mathbf{C}[f] = 0$), then (\ref{eq: Lcov}) shows simply that $na^3$ is constant in time.  

However, the evolution of the DM number density is non-trivial if the collision term exists.  To see this explicitly, consider interactions of the form $1 +  2 \leftrightarrow 3 + 4$~\cite{Gondolo:1990dk}.  The collision term for particle 1 is then
\begin{eqnarray}
g_1 \int C[f_1] \, \frac{d^3 p_1}{(2 \pi)^3} &=& -\sum_\text{spins} \int \left[ f_1 f_2 (1 \pm f_3) (1\pm f_4) |\mathcal{M}_{12\rightarrow34}|^2- f_3 f_4 (1\pm f_1) (1\pm f_2) |\mathcal{M}_{34\rightarrow12}|^2 \right] \nonumber \\ 
&&  \times (2\pi)^4 \delta^4(p_1 + p_2 -p_3- p_4) \, d \Pi_1 \,  d \Pi_2 \, d \Pi_3 \, d \Pi_4 \, ,
\label{eq: coll1}
\end{eqnarray}
where $g_i$ and $f_i$ are the spin degrees of freedom and phase-space densities, respectively, for particle  $i$, and $\mathcal{M}_{x\rightarrow y}$ is the matrix element for the reaction $x\rightarrow y$.  Factors of the form $(1\pm f)$ represent Pauli blocking and Bose enhancement; the minus sign applies to fermions and the plus sign to bosons.  These terms encapsulate the fact that it is easier(harder) for a boson(fermion) to transition to a state that already contains a boson(fermion).  The last line of (\ref{eq: coll1}) includes a delta function that enforces energy and momentum conservation, and the phase-space integration factors 
\be
d\Pi_i =  \frac{d^3 p_i}{(2\pi)^3 \, 2 E_i}. \nonumber  
\ee

In its current form, (\ref{eq: coll1}) is quite complicated; however, it reduces to a more manageable form after making the following assumptions:
\begin{enumerate}
\item \emph{Kinetic equilibrium} is maintained and so the phase-space distributions take on the Fermi-Dirac or Bose-Einstein forms.  
\item The temperature of each species satisfies $T_i \ll E_i - \mu_i$, where $\mu_i$ is its chemical potential, so that they follow the Maxwell-Boltzmann distribution.  In this case, the statistical mechanical factors in the calculation can be ignored and $(1\pm f) \sim 1$. 
\item The Standard Model particles in the interaction are in thermal equilibrium with the photon bath.  
\end{enumerate}
Using the standard definition relating the cross section to the matrix element, we get
\begin{eqnarray}
\sum_\text{spins} \int |\mathcal{M}_{ij\rightarrow kl}|^2 \times (2\pi)^4 \delta^4(p_i + p_j -p_k- p_l) \, d \Pi_k \, d \Pi_l \,  = 4 \, g_i g_j  \sigma_{ij} \sqrt{(p_i \cdot p_j)^2 - (m_i m_j)^2} \, ,\nonumber 
\end{eqnarray}
where $\sigma_{ij}$ is the cross section for the scattering process.  Substituting this back into the collision term gives  
\be
g_1 \int C[f_1] \, \frac{d^3 p_1}{(2 \pi)^3} = -\int \left\{ \left(\sigma v_\text{M\o l}\right)_{12} dn_1 dn_2 - \left(\sigma v_\text{M\o l}\right)_{34} dn_3 dn_4 \right\} \, , \nonumber
\ee
where the M\o ller velocity is defined as 
\be
\left(v_\text{M\o l} \right)_{ij}= \frac{\sqrt{(p_i \cdot p_j)^2 - (m_i \, m_j)^2}}{E_i \, E_j} \,  \nonumber
\ee
for the $ij\rightarrow kl$ process.  Because $\sigma v_\text{M\o l}$ varies slowly with changes in the number density of the initial and final-state particles, it can be factored out of the integrand to give
\be
\dot{n}_1 + 3 H n_1 = - \langle \sigma v_\text{M\o l} \rangle_{12} n_1 n_2 + \langle \sigma v_\text{M\o l} \rangle_{34}  n_3 n_4  \, . 
\label{eq: Boltz4}
\ee
Note that the velocity that is used in the cross-section average is \emph{not} the relative velocity, $v_\text{rel}$, of the incoming particles.  This is important, as $\left(v_\text{M\o l}\right)_{ij} n_i n_j$ is Lorentz invariant, whereas $ v_\text{rel} n_i n_j$ is not.  From this point forward, we will simply write the M\o ller velocity as $\vmoll \rightarrow v$ to simplify notation.

Let us now return to the specific inelastic process illustrated in Fig.~\ref{fig: equil}.  In this case, particles 1 and 2 are identical with number density $n$, and particles 3 and 4 are Standard Model particles in thermal equilibrium with the photon bath.  When the DM is also in equilibrium with the Standard Model final states, then detailed balance dictates that  
\be
\langle \sigma v \rangle_{12} \, n_\text{eq}^2 = \langle \sigma v\rangle_{34} \, n_3^\text{eq} n_4^\text{eq}  \,,   \nonumber
\ee
which can be used to rewrite the second term of (\ref{eq: Boltz4}) in terms of the DM number density and the cross section for the forward reaction.  The Boltzmann equation reduces to
\be
\dot{n} + 3 H n = \langle \sigma v \rangle \left( n_\text{eq}^2 - n^2 \right) \, ,
\label{eq: BoltzN1}
\ee
where $\langle \sigma v \rangle = \langle \sigma v \rangle_{12}$.  The DM number density, $n$, decreases with the expansion of the Universe (in addition to any number-changing effects from the collision term) and it is useful to scale out this effect by defining the quantity $Y = n/s$, where $s$ is the total entropy density of the Universe.  Substituting this into (\ref{eq: BoltzN1}) and using the fact that $sa^3$ is constant to get the relation that $\dot{s} = -3 s H$, yields
\begin{equation}
\frac{dY}{dt} = \langle \sigma v \rangle s \left( Y_\text{eq}^2 - Y^2 \right) \quad \longrightarrow \quad  
\frac{dY}{dx} = - \frac{ x s \langle \sigma v \rangle}{H(m)} \left( Y^2 - Y_\text{eq}^2 \right) \, .
\label{eq: By}
\end{equation}
This equation is written in terms of the usual time variable as well as a rescaled time variable $x = m/T$, where $m$ is the mass of the DM.  Note that $dx/dt = H(x) x$, because $T \propto 1/a$ (\emph{i.e.}, the photon temperature is inversely proportional to its wavelength, which scales as $a$).  If DM production occurs during radiation domination, then $H(x) = H(m)/x^2$.  The precise definition of $H(m)$ is not necessary for our purposes here---see~\cite{KolbTurner} for further discussion.  

Let us take stock of where we stand: We have an expression that describes the evolution of $Y$ as the Universe cools.  $Y$ is the DM number density, rescaled to remove the effects of the Universe's expansion.  Therefore, the changes in $Y$ encoded in the Boltzmann equation arise purely from interactions of the DM with states that are in thermal equilibrium with the photon bath.  The evolution of $Y$ is governed by the velocity-averaged cross section:
\be
\langle \sigma v \rangle = \frac{\int \sigma v  \, dn_1^\text{eq} \, dn_2^\text{eq}}{\int d n_1^\text{eq} \, d n_2^\text{eq}} = \frac{\int \sigma v  \, e^{-E_1/T} \, e^{-E_2/T} \, d^3 p_1 \, d^3 p_2 }{\int e^{-E_1/T} \, e^{-E_2/T} \, d^3 p_1 \, d^3 p_2 } \, .
\label{eq: sigmav}
\ee
Eq. (\ref{eq: sigmav}) can be further simplified by redefinition of the integration variables~\cite{Gondolo:1990dk}, and the final result is
\be
\langle \sigma v \rangle = \frac{1}{8 m^4 T K_2^2(m/T)} \int_{4 m^2}^\infty \sigma(\tilde{s}-4 m^2) \, \sqrt{\tilde{s}} \, K_1(\sqrt{\tilde{s}}/T) \, ds \xrightarrow{\text{non-rel.}} b_0 + \frac{3}{2} b_1 x^{-1} + \cdots\,
\label{eq: SigmaSeries}
\ee
where $K_i$ are modified Bessel functions of the $i^\text{th}$ order and $\tilde{s} = 2 m^2 + 2 E_1 E_2 - 2 \textbf{p}_1\cdot \textbf{p}_2$.  The cross section can be expanded in $x$ in the non-relativistic limit with  coefficients $b_{0,1}$, as shown.  The case where $b_0$ dominates is referred to as $s$-wave annihilation.  The case where the second term dominates is called $p$-wave annihilation.

\begin{figure}[t] 
   \centering
   \includegraphics[width=4.5in]{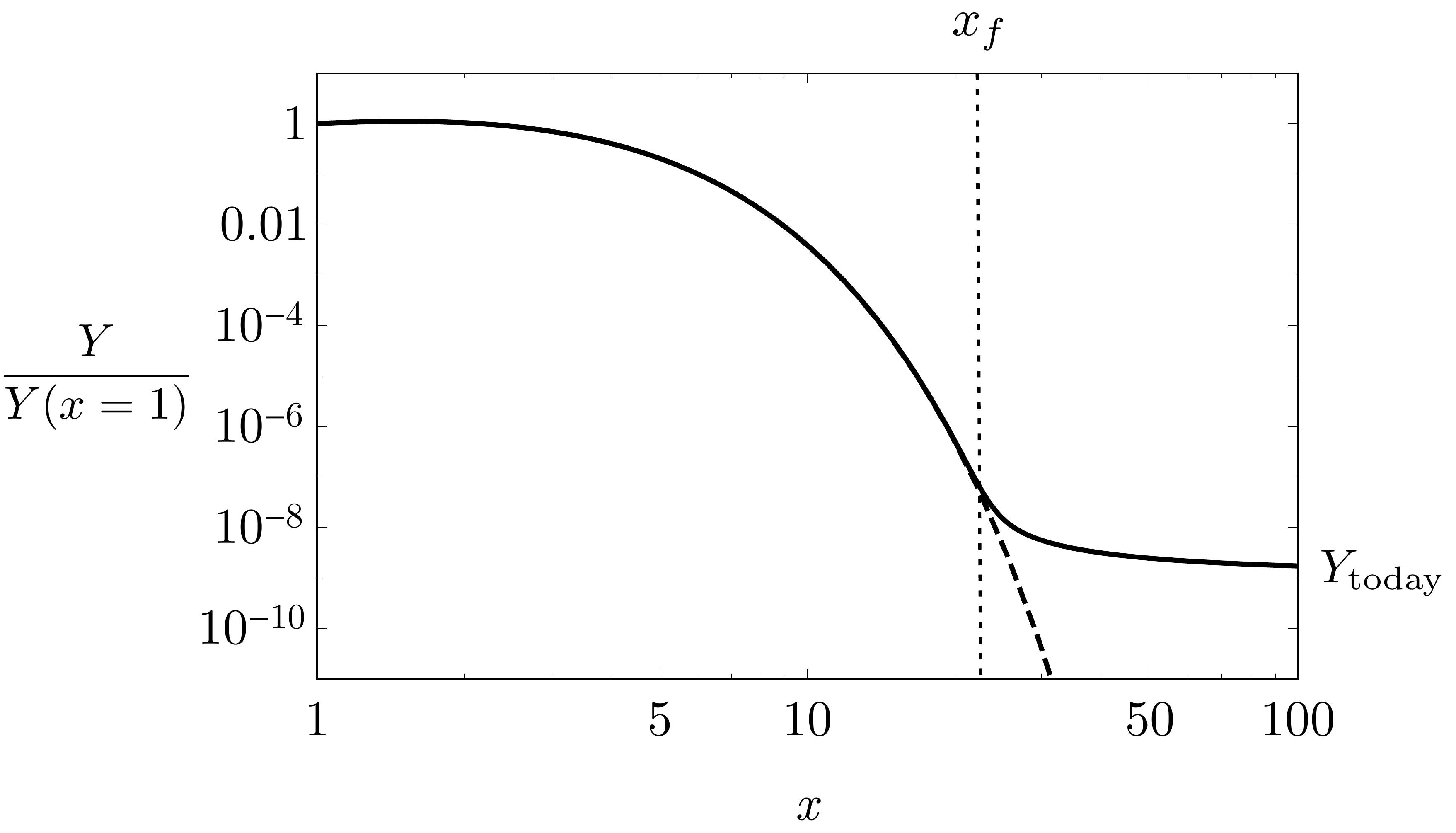} 
   \caption{An illustration of the DM number density $Y$ as a function of $x$.  Before freeze-out ($x < x_f$), the density tracks the equilibrium expectation (dashed black).  After freeze-out, the density remains nearly constant as a function of time, as indicated by the solid black line.  Figure courtesy of S.~Mishra Sharma.}
   \label{fig:freezeout}
\end{figure}

There is no analytic solution for equations that take the form of (\ref{eq: By}), so one must rely on numerical solutions for exact results.  However, we can consider the behavior of the solutions in limiting cases to build intuition for how the DM number density evolves with time.  Remember that the evolution depends on how the annihilation rate compares with the expansion rate.  When $\Gamma \gg H$, then the annihilation process is very efficient and equilibrium can be maintained between the DM and photon bath.  However, when $\Gamma \ll H$, the DM particles can no longer find each other fast enough compared to the expansion rate, and thus fall out of equilibrium, as illustrated in Fig.~\ref{fig:freezeout}.  Said another way,
\be
Y(x \lesssim x_f) \simeq Y_\text{eq}(x)  \quad \text{ and } \quad Y(x\gtrsim x_f) \simeq Y_\text{eq}(x_f) \, , \nonumber
\ee
where $x_f$ is the freeze-out time.  For CDM, $Y(x)$ decreases exponentially before freeze-out.  After freeze-out, however, the abundance is larger than what its equilibrium value would have been if freeze-out had not occurred (as $Y_\text{eq}$ is decreasing, $Y_\text{eq}(x_f) > Y_\text{eq}(x>x_f)$ trivially).  Therefore, (\ref{eq: By}) becomes
\begin{equation}
\frac{dY}{dx} \simeq - \frac{\lambda}{x^{n+2}} Y^2 \,\text{, where} \quad  \lambda = \frac{\langle \sigma v \rangle_0 s_0}{H(m)} \, . \nonumber
\end{equation}
Note that the $x$ dependence has been pulled out of the cross section and entropy to define $\lambda$.  That is, $\langle \sigma v \rangle = \langle \sigma v \rangle_0 \, x^{-n}$ and $s = s_0 \, x^{-3}$.\footnote{We are assuming that either $s$- or $p$-wave annihilation dominate, which is oftentimes true.  More precisely, though, the thermally averaged cross section is a series in $x$, as shown in (\ref{eq: SigmaSeries}). }  Taking $n=0$ as an example, we can solve for the DM abundance today: 
\be
\frac{1}{Y_\text{today}} - \frac{1}{Y_f} = \frac{\lambda}{x_f} \longrightarrow Y_\text{today} \simeq \frac{x_f}{\lambda}  \, ,  \nonumber
\ee
where the last step uses the fact that the abundance at freeze-out, $Y_f$, is typically greater than its value today.  Of course, this result changes if the thermally averaged cross-section carries a dependence on $x$, which depends on the details of the particle physics model.  If $n\neq0$, then $Y_\text{today}$ carries higher powers of $x_f$.

\vspace{0.2in}
\hrule
\vspace{0.1in}
\noindent \textsc{Exercise:} Use the fact that $n^\text{eq} \langle \sigma v \rangle \sim H$ at freeze-out to estimate that  $x_f = \mathcal{O}(10)$.  
\vspace{0.1in}
\hrule
\vspace{0.2in}

The fraction of the critical density, $\rho_\text{cr}$, contributed by the DM today is\footnote{To emphasize, (\ref{eq: Omega}) is an approximation and the full calculation gives $\Omega_\chi h^2 \propto \left[ \int_{x_f}^\infty  \langle \sigma v \rangle/x^2 dx \right]^{-1}$.  In certain cases, such as Forbidden DM, it is crucial that one properly integrate over the cross section.} 
\begin{equation}
\Omega_\chi = \frac{ m \, s_\text{today} \, Y_\text{today}}{\rho_\text{cr}} \longrightarrow \Omega_\chi h^2 \sim \frac{10^{-26} \text{ cm}^3/\text{s}}{\langle \sigma v \rangle} \simeq 0.1 \left( \frac{0.01}{\alpha} \right)^2 \left(\frac{m}{100~\text{GeV}}\right)^2 
\label{eq: Omega}
\end{equation}
taking $x_f \sim 10$ and $\langle \sigma v \rangle \sim \alpha^2/m^2$.  Assuming a weakly interacting DM particle with $\alpha \sim 0.01$ and mass $m_\chi \sim 100$~GeV gives the correct abundance today as measured by Planck and WMAP~\cite{Ade:2015xua}.  The fact that weak-scale DM naturally gives the correct DM density today is known as the `WIMP miracle' and has become the dominant paradigm as many well-motivated models, such as supersymmetry, provide such candidates.\footnote{In writing (\ref{eq: Omega}), we have assumed that the DM annihilation is a 2$\rightarrow$2 process.  Because unitarity prevents the cross section from being arbitrarily large, it is possible to set an upper limit on the mass of thermal DM to be $m_\chi \lesssim 100 \text{ TeV}$~\cite{Griest:1989wd}.  This bound is model-dependent, though, in that it assumes that no additional particles exist that are heavier than the DM.}  Such particles are known as Weakly Interacting Massive Particles, or WIMPs.  As we will discuss, there is a wide-ranging experimental program today targeting this parameter space.  

But just how much of a miracle are WIMPs?  It turns out, not that much.  Going back to (\ref{eq: Omega}), notice that what is really constrained is the \emph{ratio} of the squared coupling to the mass.  Indeed, it is possible to open up a wider band of allowed masses for thermal DM by taking $\alpha \ll 1$ while keeping $\alpha^2/m^2$ fixed.  Such scenarios are known as WIMPless DM models~\cite{Feng:2008ya}.  

A separate example that easily generates DM with masses down to the keV scale is known as Forbidden DM~\cite{Griest:1990kh, D'Agnolo:2015koa}.  Forbidden DM arises when the DM annihilates primarily into some new, heavier particles 
\be
\chi \chi \rightarrow \phi \phi \, , \noindent \nonumber
\ee
with $m_{\phi} > m_\chi$.  Note that the $\phi$ are not Standard-Model states, but we do assume that they are in equilibrium with the photon bath during freeze-out.   In this case, the Boltzmann equation becomes
\be
\dot{n} + 3 H n = - \left \langle \sigma v \right\rangle_{\chi \chi} n_\chi^2  + \left \langle \sigma v \right \rangle_{\phi \phi} (n^\text{eq}_{\phi} )^2 \, \nonumber
\label{eq: BoltzN}
\ee
and we proceed, as above, by rewriting this in terms of $n_\chi$ and $\langle \sigma v \rangle_{\chi \chi}$, and solving for the DM density today.  At this point, however, we need to be more careful.  When the DM annihilates to heavier final states, we cannot simply approximate the thermally averaged cross section $\langle \sigma v \rangle_{\chi \chi}$ as $\alpha^2/m_\chi^2$ because important phase-space suppression factors come into play.  The reverse reaction, however, does scale as $\langle \sigma v \rangle_{\phi \phi} \sim \alpha^2/m_\phi^2$.  To relate it to $\langle \sigma v \rangle_{\chi \chi} $,  we take advantage of the following relation from detailed balance:
\be
\left \langle \sigma v \right\rangle_{\chi \chi}  = \left \langle \sigma v \right \rangle_{\phi \phi} \left(\frac{n^\text{eq}_{\phi}}{n^\text{eq}_\chi}\right)^2 \,  \sim \frac{\alpha^2}{m_\phi^2} \, e^{-(m_\phi-m_\chi)/T} \, .  \nonumber
\ee 
In the interesting regime when the mass difference is small but non-zero,  the thermally averaged cross section for the forward reaction is exponentially suppressed.  This provides the freedom to reduce the DM mass while still keeping the DM density at its target value (while not changing the coupling too far from weak-scale).  Forbidden DM provides a simple illustration of how a (slightly) more complicated DM model can give the correct relic density for masses below the weak scale.  The Forbidden example still uses $2\rightarrow2$ DM interactions.  Broadening this assumption can lead to intriguing consequences as well.  For example,  if the relic density is set by $3\rightarrow2$ interactions, then a strongly-interacting MeV-scale thermal relic is allowed~\cite{Carlson:1992fn,Hochberg:2014dra}.   

\section{Application 1: Direct Detection}

Our primary focus thus far has been understanding the starting assumptions for the astrophysical and particle-physics properties of DM.  Now, we turn to several applications of these results, focusing specifically on detection strategies for WIMPs.  This section will review the physics of direct detection experiments, and the next section will focus on indirect detection.    

Imagine DM flying through Earth and scattering off a particle in a ground-based detector, which then recoils with some energy $E_R$.  If the recoil energy is large enough, it may be possible to detect the scattered particle and infer, from its kinematics, the properties of the DM that scattered off it.  The `direct detection' of DM in this fashion was first proposed by Goodman and Witten~\cite{Goodman:1984dc} and then developed more fully by Drukier, Freese, and Spergel~\cite{Drukier:1986tm} in the mid-1980s.     

If the DM scatters off a nucleus with mass $m_N$, then the nuclear recoil energy is 
\be
E_R = \frac{q^2}{2 m_N} \simeq 50~\text{keV} \left( \frac{m_\chi}{100~\text{GeV}} \right)^2 \, \left( \frac{100~\text{GeV}}{m_N} \right)\, , 
\label{eq: ER}
\ee
where $q\sim m_\chi v$ is the momentum transfer in the collision and $v\sim10^{-3}$ is the (non-relativistic) speed of the incoming DM.\footnote{For sub-GeV DM, scattering off electrons in the target, rather than nuclei, is more relevant.  In this case, evaluation of the scattering rate is more involved as one must account for the fact that the electron is in a bound state.  We point the interested reader to~\cite{Essig:2011nj,Essig:2015cda,Lee:2015qva} for further details.}  Let us take as an example the LUX~\cite{Akerib:2015rjg} and Xenon100~\cite{Aprile:2012nq} experiments, which use a Xenon target with mass $m_N \sim 120$~GeV.  These experiments have  energy thresholds $\sim$few~keV, so (\ref{eq: ER}) tells us that their sensitivity degrades for $m_\chi\lesssim10$~GeV.  However, they are optimal for detecting DM with $m_\chi\sim 100$~GeV, where $E_R$ is on the order of $\sim$tens of keV.  

The kinetic energy of an incident DM particle with mass of $100$~GeV is $\sim$10~keV, which is much smaller than the order 1--10~MeV nuclear binding energy of an atomic target.  As a result, we need only consider the scattering of the DM off the nucleus as a whole (as opposed to its constituents).  

\subsection{Scattering Rate}

The basic quantity of interest is the scattering rate of the DM particle off the nuclear target.  The differential rate per unit detector mass is
\begin{equation}
\frac{dR}{dE_R} = \frac{n_\chi}{m_N} \left \langle v \, \frac{d\sigma}{dE_R} \right \rangle \, , \nonumber
\end{equation}
where $n_\chi = \rho_\chi/m_\chi$ is the DM number density and $d\sigma/dE_R$ is the differential scattering cross section.  The brackets indicate an average over the velocities of the incoming DM.  Written out in full, the differential rate is 
\be
\frac{dR}{dE_R} =  \frac{\rho_\chi}{m_\chi \, m_N} \int_{v_\text{min}}^{v_\text{max}} d^3 v \,\, v \, \tilde{f}\left(\mathbf{v}, t\right) \,\frac{d\sigma}{dE_R}\, , 
\label{eq: dRdER}
\ee
where $\tilde{f}\left( \mathbf{v}, t \right)$ is the DM velocity distribution in the lab frame, $v_\text{max}$ is the escape velocity, and $v_\text{min}$ is the minimum velocity needed to cause a nucleus to scatter with energy $E_R$.  Measurements of the fastest stars in the Galaxy bound the escape velocity to be within 498--608~km/s, to 90\% confidence~\cite{Smith:2006ym}.  

The lab-frame velocity distribution is obtained by applying a Galilean boost to the Galactic-frame distribution, $f(\mathbf{v})$:
\be
\tilde{f}\left(\mathbf{v}\right) = f\left( \mathbf{v} +  \mathbf{v}_\text{obs}\left(t\right)  \right) \, , \text{ where } \mathbf{v}_\text{obs}\left(t\right) = \mathbf{v}_\odot + \mathbf{V}_\oplus\left(t\right) \, , \nonumber
\ee
$\mathbf{v}_\odot$ is the velocity of the Sun relative to the DM reference frame~\cite{Schoenrich:2009bx, Mignard}, and $\mathbf{V}_\oplus(t)$ is the velocity of the Earth about the Sun.\footnote{There are errors in the formula of the Earth's velocity scattered throughout the literature.  For the updated value, see~\cite{Lee:2013xxa}.}  For reference, $v_\odot \sim 220$~km/s and $V_\oplus \sim 30$~km/s.  To good approximation, 
\be
\mathbf{v}_\text{obs}(t) \approx \mathbf{v}_\odot \left( 1+ \epsilon \cos\left[ \omega (t-t_0) \right] + \cdots \right) \, \nonumber
\ee
where $\omega = 2\pi/\text{year}$, $t_0$ is the phase of the modulation, and $\epsilon$ is the ratio of $\widetilde{V}_\oplus/v_\odot$.  Here, $\widetilde{V}_\oplus$ is the component of the Earth's velocity in the Sun's direction.  Because $\epsilon \ll 1$, the velocity distribution can be Taylor expanded as 
\be
f\left(\mathbf{v} + \mathbf{v}_\text{obs} (t) \right) \simeq  f(\textbf{v} + \textbf{v}_\odot ) + \epsilon \cos\left[ \omega ( t - t_0) \right] f'(\textbf{v} + \textbf{v}_\odot) + \cdots \, \nonumber
\ee
and the rate equation takes the form
\be
\frac{dR}{dE_\text{nr}} = A_0 + A_1\cos\left[\omega \, (t-t_0)\right] + \cdots  \, .
\label{eq: taylor}
\ee
The first term ($A_0$) is the unmodulated rate and the second term ($A_1$) describes the annual modulation of the signal, which we will come back to later.  The higher order terms in the expansion may be relevant in cases where the DM is light ($\lesssim 10$~GeV) or in the presence of velocity substructure~\cite{Lee:2013xxa}, but we will not discuss them here.  

Calculating the rate requires knowing $v_\text{min}$, which depends on the kinematics of the scattering event.  We consider the general case $\chi + \text{N} \rightarrow \chi' + \text{N}$\,, where $\chi'$ is an excited state of the DM particle with mass $m_\chi + \delta$.  Note that the elastic scattering regime is recovered when $\delta \rightarrow 0$.  In some models, the inelastic process may dominate, so we will work out the general form for $v_\text{min}$ here for reference.  In the non-relativistic limit, the initial momenta in the lab frame are
\be
p_\mu = \left( m_N , \mathbf{0} \right) \quad \text{ and } \quad k_\mu = \left( m_\chi + \frac{1}{2} m_\chi v^2, m_\chi v, 0, 0 \right) \\ \nonumber
\ee
for the nucleus and DM, respectively.  (The corresponding momenta for the final states are denoted with primes.)  It is often more convenient to use the momentum transfer $\mathbf{q}$ and the total momentum $\mathbf{P}$,
\be
\mathbf{q} = \mathbf{p}' - \mathbf{p} = \mathbf{k} - \mathbf{k}' \quad \text{ and } \quad 
\mathbf{P} = \mathbf{p}' + \mathbf{p} = \mathbf{q} + 2 \mathbf{p} \, ,  \nonumber
\ee
and to define
\be
q_\mu = \left(E_R, \sqrt{2 m_N E_R} \cos\theta, \sqrt{2 m_N E_R} \sin\theta, 0\right) \, , \nonumber
\ee
where $\theta$ is the scattering angle of the nucleus in the lab frame.  

\vspace{0.2in}
\vbox{
\hrule
\vspace{0.1in}
\noindent \textsc{Exercise:} Show that 
\be
q\cdot(k-p) - q^2 = E_R \, (m_\chi + m_N) - m_\chi v \, \sqrt{2m_N E_R} \cos\theta =  -m_\chi \, \delta \, .
\label{eq: midstep}
\ee
\hrule
}
\vspace{0.2in}

\noindent From (\ref{eq: midstep}), we solve for $v \cos\theta$ and find that\footnote{Note that $\delta$ can be negative in the case of an exothermic interaction~\cite{Graham:2010ca, Essig:2010ye}.} 
\be
v_\text{min} = \frac{1}{\sqrt{2 m_N E_R}} \left| \frac{E_R \, (m_\chi + m_N)}{m_\chi} + \delta \right| \, .  \nonumber
\ee
Our discussion will focus primarily on the elastic-scattering regime where
\be
v_\text{min} = \sqrt{\frac{m_N \, E_R}{2 \, \mu^2}} \,   \nonumber
\ee
and $\mu$ is the reduced mass of the DM-nucleus system.  Notice that the minimum velocity to scatter increases(decreases) as $m_N$($m_\chi$) increases, as our intuition would suggest.  Additionally, $v_\text{min}$ is larger for inelastic scattering events.  

\begin{figure}[tb] 
   \centering
   \includegraphics[width=6in]{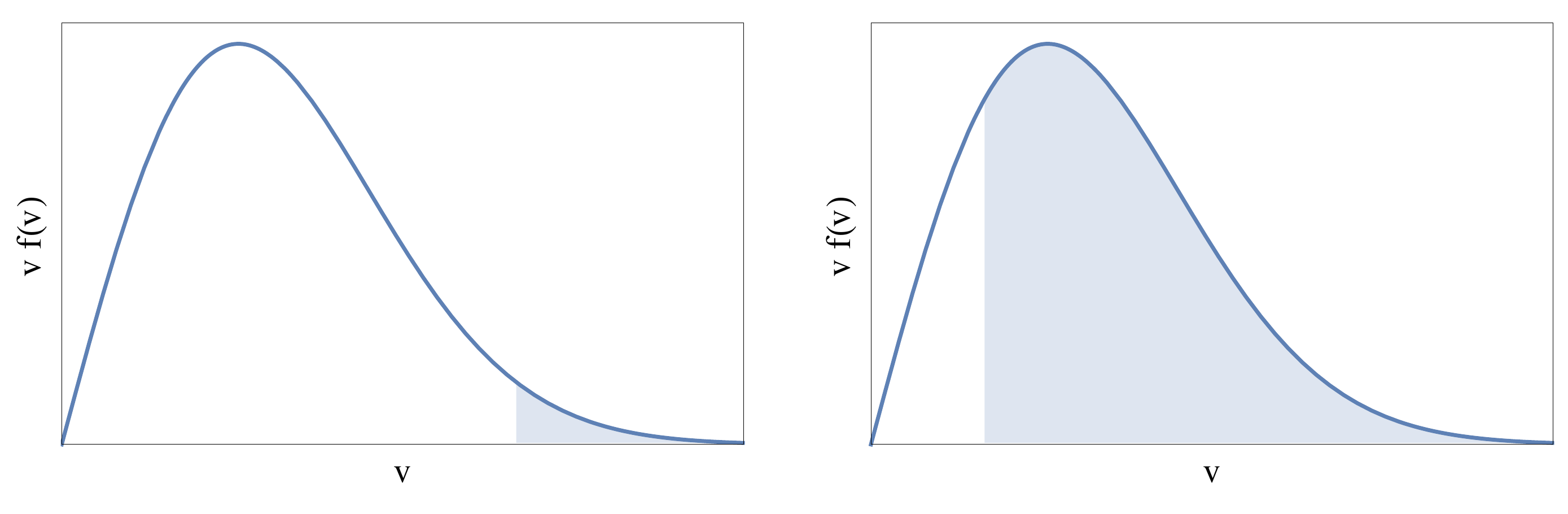} 
   \caption{Schematic illustration of the integration region for a lighter DM candidate (left) versus a heavier candidate (right), scattering off the same nucleus. }
   \label{fig: vrange}
\end{figure}

Taking the Standard Halo Model and assuming that $d\sigma/dE_R \propto 1/v^2$ (as we will motivate shortly), then the unmodulated rate is approximately
\be
\frac{dR}{dE_R} \propto \int_{v_\text{min}}^{v_\text{esc}} d^3 v \,\, \frac{f(\mathbf{v} + \mathbf{v}_\odot)}{v} \sim \int_{v_\text{min}}^{v_\text{esc}} dv \, v \, e^{-v^2/v_0^2} \sim e^{-E_R/E_0} \, , \nonumber
\label{eq: diffRapprox}
\ee
where $E_0 = 2 \mu^2 v_0^2/m_N$.\footnote{Those interested in techniques for finding exact solutions to these integrals should refer to the classic review article~\cite{LewinSmith}.}  For a 100~GeV DM scattering off a Xenon target, $E_0\sim 50$~keV.  This means that the expected recoil spectrum for the nucleus is exponentially falling, for typical assumptions about the cross section and velocity distribution.  Figure~\ref{fig: vrange} shows a sketch of the integration region for a given target.  The left(right) panels are examples for lower(higher) DM mass.  Clearly, as the DM mass increases, the rate is larger as the integration region is larger.  Notice that in the case of light DM, when $v_\text{min}$ is large, the rate becomes very sensitive to the tail of the velocity distribution, which, as we already discussed, can be quite uncertain.  

\vspace{0.2in}
\hrule
\vspace{0.1in}
\noindent \textsc{Exercise:} Sketch the differential scattering rate as a function of nuclear recoil energy for inelastic DM versus elastic DM.  
\vspace{0.1in}
\hrule
\vspace{0.2in}

\subsection{Differential Scattering Cross Section}

Now, we turn to deriving the differential scattering cross section for the DM-nucleus interaction, taking an effective operator approach~\cite{Fan:2010gt,Fitzpatrick:2012ix}.  Let us assume that the DM is a spin-1/2 Dirac fermion    
that interacts with quarks via a scalar or vector boson $\phi$ with mass $m_\phi$.  The scattering process is described by the effective four-fermion interaction:
\be
\mathcal{L}_\text{eff} = g(q^2, m_\phi) \, \bar{\chi} \, \Gamma_\chi \, \chi  \,\, \bar{Q}  \, \Gamma_Q  \, Q \, ,  \nonumber
\ee
where $Q$ represents the quark fields, $\Gamma_{\chi, Q} = \{I, \gamma^5, \gamma^\mu, \gamma^\mu \gamma^5, \sigma^{\mu \nu}, \sigma^{\mu \nu} \gamma^5$\}, and $g(q^2, m_\phi)$ is an effective coupling.  In particular, $g(q^2, m_\phi)$ is proportional to $1/m^2_\phi$ for contact interactions ($q^2 \ll m_\phi^2$), or $1/q^2$ for long-range interactions ($q^2 \gg m_\phi^2$).  We then proceed as follows:     
\begin{enumerate}
\item Map the quark operator to a nucleon operator and use this to obtain the amplitude for DM-nucleus scattering, $\mathcal{M}$.
\item Take the non-relativistic limit of the scattering amplitude, $\mathcal{M}_\text{nr}$.     
\item Relate this to the differential cross section by averaging/summing initial and final-state spins:
\be
\frac{d\sigma}{dE_R} = \frac{2 m_N}{ \pi v^2} \left\langle \left| \mathcal{M}_\text{nr} \right|^2 \right\rangle \, , \nonumber  
\ee
where $v$ is the relative velocity between the DM and nucleus. 
\end{enumerate}

\noindent As a first example, consider the effective Lagrangian 
\be
\mathcal{L}_\text{eff} = g_\phi \, \bar{\chi} \, \chi \, \bar{Q} \, Q  
\label{eq: Leff1}
\ee
for contact interactions where $g_\phi$ is independent of the momentum transfer.  To rewrite the quark fields in terms of nucleon fields (labeled $n$, $p$), we must evaluate operators of the form $\left\langle n \left| \bar{Q} Q \right| n \right\rangle$.  These terms are related to the nucleon mass using the trace of the QCD energy-momentum tensor---for further details, see~\cite{Jungman:1995df}.  The fraction of the proton mass accounted for by a particular quark flavor is defined as $m_p f_{T_q}^p \equiv \left \langle p \left| m_q \bar{Q} Q \right| p \right\rangle$, and the coupling of the DM to the protons is given by
\be
f_p = \sum_{q=u,d,s} m_p \frac{g_\phi}{m_q} \, f_{T_q}^p + \frac{2}{27} f_{T_G}^p \sum_{q=c,b,t} m_p \frac{g_\phi}{m_q} \,  ,
\label{eq: fp}
\ee
where $f_{T_G}^p = 1- \sum_{q=u,d,s} f_{T_q}^p$~\cite{Cerdeno:2010jj}.  A similar relation applies to the DM coupling to neutrons, with appropriate substitutions.  The mass fractions $f_{T_q}^p$ are determined experimentally, so $f_{p,n}$ are constants of the theory once $g_\phi$ is set.  The scattering amplitude is therefore
\be
\mathcal{M} = f_p \, \bar{\chi} \,  \chi \, \bar{p} \, p + f_n \, \bar{\chi} \, \chi \, \bar{n} \, n \, . \nonumber
\ee
Because the momentum transfer is small enough that the nucleon structure cannot be resolved, no form factors need to be included here.  In many models, the DM couples to protons and neutrons with the same strength, so that $f_p \approx f_n$.  However, isospin-dependent scenarios have also been considered in the literature~\cite{Feng:2011vu}.  

Because $\bar{p}p$ and $\bar{n} n$ give the proton and neutron count, respectively, it is straightforward to rewrite $\mathcal{M}$ in terms of the fields for the nuclei: 
\be
\mathcal{M} = \left[ Z f_p + (A-Z) f_n \right] \, \bar{\chi} \,  \chi \, \bar{N} \, \Gamma_N \, N  \, , \nonumber
\ee
where $Z$ is the atomic number, $A$ is the mass number, and $\Gamma_N$ is a Lorentz-invariant 4$\times$4 matrix.  Because the final answer can only depend on $q_\mu$ and $P_\mu$, there are a limited number of possibilities for $\Gamma_N$:
\be
\bar{N} \Gamma_N N = \bar{N} N \, \tilde{F}_1(q) + \bar{N} \gamma^\mu N \, q_\mu \, \tilde{F}_2(q) + \bar{N} \gamma^\mu N \, P_\mu \,  \tilde{F}_3(q) + \bar{N} \sigma^{\mu \nu} N \, q_\mu P_\nu \, \tilde{F}_4(q)  \, ,
\label{eq: ScalNuc}
\ee
where $\tilde{F}_i(q)$ are nuclear form factors.  At small momentum transfers, the DM does not probe the size of the nucleus and the cross section is unaffected.  However, as the momentum transfer increases, the interactions become sensitive to the size of the nucleus and the cross section is diminished.  This effect is encoded in the form factors.  
The Dirac equation tells us that $\gamma^\mu p_\mu N(p) = m_N N(p)$ and $\bar{N}(p') \gamma^\mu p'_\mu = m_N \bar{N}(p')$, which means that the second term in (\ref{eq: ScalNuc}) vanishes, while all others are proportional to $\bar{N}N$.  Therefore, 
\be
\mathcal{M} = \left[ Z f_p + (A-Z) f_n \right] \, \bar{\chi} \,  \chi \, \bar{N} \, N  \, F(q) \, , \label{eq: phiM}
\ee
where $F(q)$ is a linear combination of the $\tilde{F}_i$'s.
For interactions such as this one that are coherent over the entire nucleus, the form factor is approximately the Fourier transform of the nucleus' mass distribution.  It is commonly given by the Helm form factor~\cite{Helm:1956zz}:
\be
F(q) = 3 e^{-q^2 s^2/2} \, \frac{\sin (q\,r_n) - q\,r_n \cos(q\,r_n)}{(q\,r_n)^3} \, , \nonumber
\ee
where the effective nuclear radius is $r_n^2 = c^2 + \frac{7}{3} \pi^2 a^2 - 5s^2$, with $a\simeq 0.52$~fm, $s\simeq 0.9$~fm, and $c\simeq 1.23A^{1/3} - 0.60$~fm.  For a detailed discussion on the form factor, see~\cite{LewinSmith}.  

Next, we want to find the non-relativistic limit of the amplitude (\ref{eq: phiM}).  Remember that Dirac fields are given by 
\[ N^s(p) = \left(\begin{array}{c}
\sqrt{p \cdot \sigma} \, \xi^s  \\
\sqrt{p \cdot \bar{\sigma}} \, \xi^s \end{array}\right) \, ,\] 
where $s$ is the spin index and $\xi^s$ is the two-component spinor satisfying $\sum_\text{spins}\xi^{s\dagger} \xi^s = 1$.  In the non-relativistic limit, $p^0 \approx m_N$ and 
\be
\sqrt{p\cdot \sigma} \approx \sqrt{ m_N - \mathbf{p} \cdot \mathbf{\sigma} } \approx \sqrt{m_N} \left( 1- \frac{  \mathbf{p} \cdot \mathbf{\sigma}}{2 m_N} \right) \, . \nonumber
\ee
The same applies for the $\chi$ fields, except with appropriate substitutions for mass and momenta.  Therefore,
\begin{eqnarray}
\nonumber
\bar{N}^{s'}(p') N^s(p) &=& (N^{s'}(p'))^\dagger \gamma^0 N^s(p) \\ \nonumber
&=& \left( \sqrt{p'\cdot \sigma} \,\xi^{s' \dagger} \, , \sqrt{p' \cdot \bar{\sigma}} \, \xi^{s' \dagger} \right)
\left(\begin{array}{cc}
0 & 1  \\
1 & 0 \end{array}\right)
\left(\begin{array}{c}
\sqrt{p \cdot \sigma} \, \xi^s  \\
\sqrt{p \cdot \bar{\sigma}} \, \xi^s  \end{array}\right) \\ \nonumber
&=& \xi^{s' \dagger} \left(\sqrt{p' \cdot \bar{\sigma}} \sqrt{p\cdot \sigma} + \sqrt{p' \cdot \sigma} \sqrt{p\cdot \bar{\sigma}} \right) \xi^s \\ \nonumber
&\approx& 2 m_N \,  \xi^{s' \dagger} \xi^s\nonumber \, ,
\end{eqnarray} 
where $s(s')$ is the spin index for the incoming(outgoing) nucleus.  Similarly, $\bar{\chi} \chi \approx 2 m_\chi \, \xi^{r' \dagger} \xi^r$ in the non-relativistic limit, where $r(r')$ is the spin index for the incoming(outgoing) DM.  Dropping the factors of $2 m_\chi$ and $2 m_N$, which are relativistic normalizations, gives
\be
\mathcal{M}_\text{nr} = \left[ Z f_p + (A-Z) f_n \right] \, F(q) \, \xi^{s' \dagger} \xi^s \,  \xi^{r' \dagger} \xi^r  \, .\nonumber
\ee
The differential scattering cross section is thus
\be
\frac{d\sigma}{dE_R} = \frac{2 m_N}{\pi v^2} \frac{1}{(2 J+1) (2s_\chi +1)} \sum_\text{spins} \left[ Z f_p + (A-Z) f_n\right]^2 \, F^2(q) \, \left|  \xi^{s' \dagger} \xi^s \right|^2 \left|  \xi^{r' \dagger} \xi^r \right|^2  \, , \nonumber
\ee
where $J(s_\chi)$ is the nuclear(DM) spin.  Note that
\be
\frac{1}{2 s_\chi +1 } \sum_{r',r = 1,2} \left|  \xi^{r' \dagger} \xi^r \right|^2 = \frac{1}{2 s_\chi +1 } \sum_{r',r} \text{tr} \left[ \xi^{r'} \xi^{r' \dagger} \xi^{ r } \xi^{r \dagger} \right] = \frac{1}{2} \text{tr} \left[1 \right] = 1 \, . \nonumber
\ee
A similar result applies to the spinor product of the $\xi^s$, leaving us with
\be
\frac{d\sigma}{dE_R} = \frac{2 m_N}{\pi v^2}  \left[ Z f_p + (A-Z) f_n\right]^2 \, F^2(q)  \, . 
\label{eq: sigmaSI}
\ee
There are a few important points to note regarding (\ref{eq: sigmaSI}).  First, when $f_p = f_n$, the differential cross section is proportional to $A^2$.  In this case, the DM couples coherently to the entire nucleus and the strength of the scattering interaction increases with the mass number of the nucleus.  Second, effective interactions such as (\ref{eq: Leff1}) are referred to as `spin-independent' because the scattering cross section does not depend on the nuclear spin.  Spin-independent interactions of the form of (\ref{eq: Leff1}) are often cited in the literature because they arise naturally in models of supersymmetry with neutralino DM.  Third, the scattering cross section is independent of the recoil energy and thus the differential rate is a falling exponential.

It is important to keep in mind that the example above is not generic and, indeed, predictions for the differential rate can change markedly by analyzing different effective interactions.  For instance, consider what happens when $g_\phi \propto 1/q^2 \propto 1/E_R$ in (\ref{eq: Leff1}).  It is convenient to factor out the momentum dependence from (\ref{eq: fp}) and introduce a DM form factor, $F_\text{DM}(E_R) \propto E_R^{-1}$, that multiplies the amplitude.  In this case, the differential cross section is proportional to $d\sigma/dE_R \propto E_R^{-2}$, and the scattering rate is enhanced at small recoil energies, relative to the case of contact interactions.

It is also possible for the recoil spectrum to be suppressed at low energies, rather than enhanced.  Consider, as an example, 
\be
\mathcal{L}_\text{eff} = g_\phi \, \bar{\chi} \, \gamma^5 \, \chi \, \bar{Q} \, Q \, .  
\nonumber
\ee
The amplitude in this case is nearly identical to what we calculated above for the scalar operator, except for 
\begin{eqnarray}
\nonumber
\bar{\chi}^{r'}(k') \gamma^5 \chi^r(k) &=& (\chi^{r'}(k'))^\dagger \gamma^0 \gamma^5 \chi^r(k) \\ \nonumber
&=& \left( \sqrt{k'\cdot \sigma} \,\xi^{r' \dagger} \, , \sqrt{k' \cdot \bar{\sigma}} \, \xi^{r' \dagger} \right)
\left(\begin{array}{cc}
0 & 1  \\
1 & 0 \end{array}\right)
\left(\begin{array}{cc}
-1 & 0  \\
0 & 1 \end{array}\right)
\left(\begin{array}{c}
\sqrt{k \cdot \sigma} \, \xi^r  \\
\sqrt{k \cdot \bar{\sigma}} \, \xi^r  \end{array}\right) \\ \nonumber
&=& \xi^{r' \dagger} \left(- \sqrt{k' \cdot \bar{\sigma}} \sqrt{k\cdot \sigma} + \sqrt{k' \cdot \sigma} \sqrt{k\cdot \bar{\sigma}} \right) \xi^r \\ \nonumber
&\approx& - \xi^{r' \dagger} \, \left( \mathbf{q} \cdot \mathbf{\sigma} \right)\, \xi^r\nonumber \, = -  \mathbf{q} \cdot \mathbf{s}_\chi,
\end{eqnarray} 
which needs to be expanded to higher order in the momentum transfer.  Therefore, the non-relativistic amplitude is proportional to $\mathbf{q}$ and the scattering amplitude goes as $q^2$, or $d\sigma/dE_R \propto E_R$.     

In other cases, the differential cross section may depend on the spin of the nucleus.  The typical example for spin-dependent interactions is 
\be
\mathcal{L}_\text{eff} \propto \bar{\chi} \, \gamma_\mu \, \gamma^5 \, \chi \, \bar{Q} \, \gamma^\mu \, \gamma^5 \, Q \, , \nonumber
\ee
with cross section 
\be
\frac{d\sigma}{dE_R} = \frac{16 m_N}{\pi v^2} G_F^2 J(J+1) \Lambda^2 F_\text{SD}^2(q)\,  ,
\label{eq: SD}
\ee
where $\Lambda \equiv \frac{1}{J} \left( a_p \left\langle S_p \right\rangle + a_n \left\langle S_n \right\rangle \right)$~\cite{Jungman:1995df}.  Here, $G_F$ is the Fermi coupling constant, $a_{p(n)}$ is the effective coupling of the DM to the proton(neutron), and $\langle S_{p(n)} \rangle$ is the average spin contribution of the proton(neutron).  Importantly, the spin-dependent form factor is different than the spin-independent form factor---see~\cite{Jungman:1995df} for further discussion.  Notice that the spin-dependent interaction is no longer coherent with the nucleus and (\ref{eq: SD}) does not scale as $A^2$.  As a result, spin-dependent interactions are more challenging to observe experimentally and the current bounds are weaker than those from spin-independent interactions.
\begin{figure}[tb] 
   \centering
   \includegraphics[width=5in]{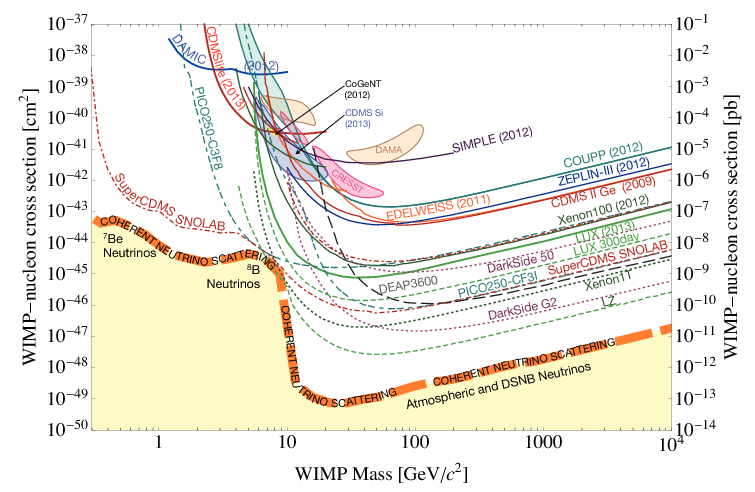} 
   \caption{Summary of current (solid) and projected (dotted/dashed) bounds on the spin-independent WIMP-nucleon cross section.  Shaded regions denote experimental anomalies, all of which are in tension with the exclusion bounds.  The thick orange line denotes the cross section below which the experiments become sensitive to coherent neutrino scattering off nuclei~\cite{Billard:2013qya}.  Figure from~\cite{Cooley:2014aya}.} 
   \label{fig: ddlimits}
\end{figure}

Figure~\ref{fig: ddlimits} is a compilation of results from current direct detection experiments (solid lines), as well as projections for future experiments (dotted/dashed lines)~\cite{Cooley:2014aya}.  The limits are plotted for the canonical spin-independent scenario with $\mathcal{L}_\text{eff} \propto \bar{\chi} \chi \bar{Q} Q$ and a heavy mediator.  Because the different experiments use a variety of target nuclei, it is not ideal to compare the DM-\emph{nucleus} cross section, $\sigma_\text{SI}$, between experiments.  Instead, we factor out the dependence on the target nucleus and use the DM-\emph{nucleon} cross section, $\sigma_p$.  The two are related by
\be
\sigma_\text{SI} = \frac{\mu^2}{\mu_p^2} A^2 \sigma_p \, , \,\, \text{ where $\sigma_\text{SI}$ is defined such that} \quad
\frac{d\sigma}{dE_R} = \frac{2 m_N}{4 \mu^2 v^2} \, \sigma_\text{SI} \, F^2(q) 
 \nonumber
\ee
with $d\sigma/dE_R$ given in (\ref{eq: sigmaSI}) and $\mu$($\mu_p$) the reduced mass of the DM and nucleus(nucleon).  Figure~\ref{fig: ddlimits} plots the limits in terms of $\sigma_p$ and the DM mass.  Notice that the bounds become weaker at masses $m_\chi \lesssim 10$~GeV due to the energy thresholds of the experiments.  Across all experiments, the sensitivity is optimal $\sim$50--100~GeV, and then weakens towards higher DM mass because the DM number density scales as $1/ m_\chi$.  The most sensitive experiments are currently starting to probe DM-nucleon cross sections $\sim$10$^{-45}$~cm$^2$, which is in the range expected for DM that interacts with the nucleus via the exchange of a Higgs boson.  

\vspace{0.2in}
\vbox{
\hrule
\vspace{0.1in}
\noindent \textsc{Exercise:} Show that Higgs exchange between the DM and nucleus is a spin-independent interaction and estimate the magnitude of the scattering cross section.
\vspace{0.1in}
\hrule
}
\vspace{0.2in}

\noindent Below the thick orange band in Fig.~\ref{fig: ddlimits}, coherent neutrino scattering becomes relevant and the experiments are no longer background-free~\cite{Billard:2013qya}.  The shaded regions in Fig.~\ref{fig: ddlimits} correspond to detections of excess events.  It is challenging to interpret these as DM detections given that other experiments simultaneously exclude the same regions of parameter space, however we will not delve into this debate here.  Note that limit plots such as Fig.~\ref{fig: ddlimits} usually assume the Standard Halo Model and the results can look different if the velocity distribution is varied.  It is feasible to present limits independent of astrophysical assumptions---see  \emph{e.g.}, \cite{Fox:2010bu,Fox:2010bz,Fox:2014kua}.  

\begin{figure}[tb] 
   \centering
   \includegraphics[width=3in]{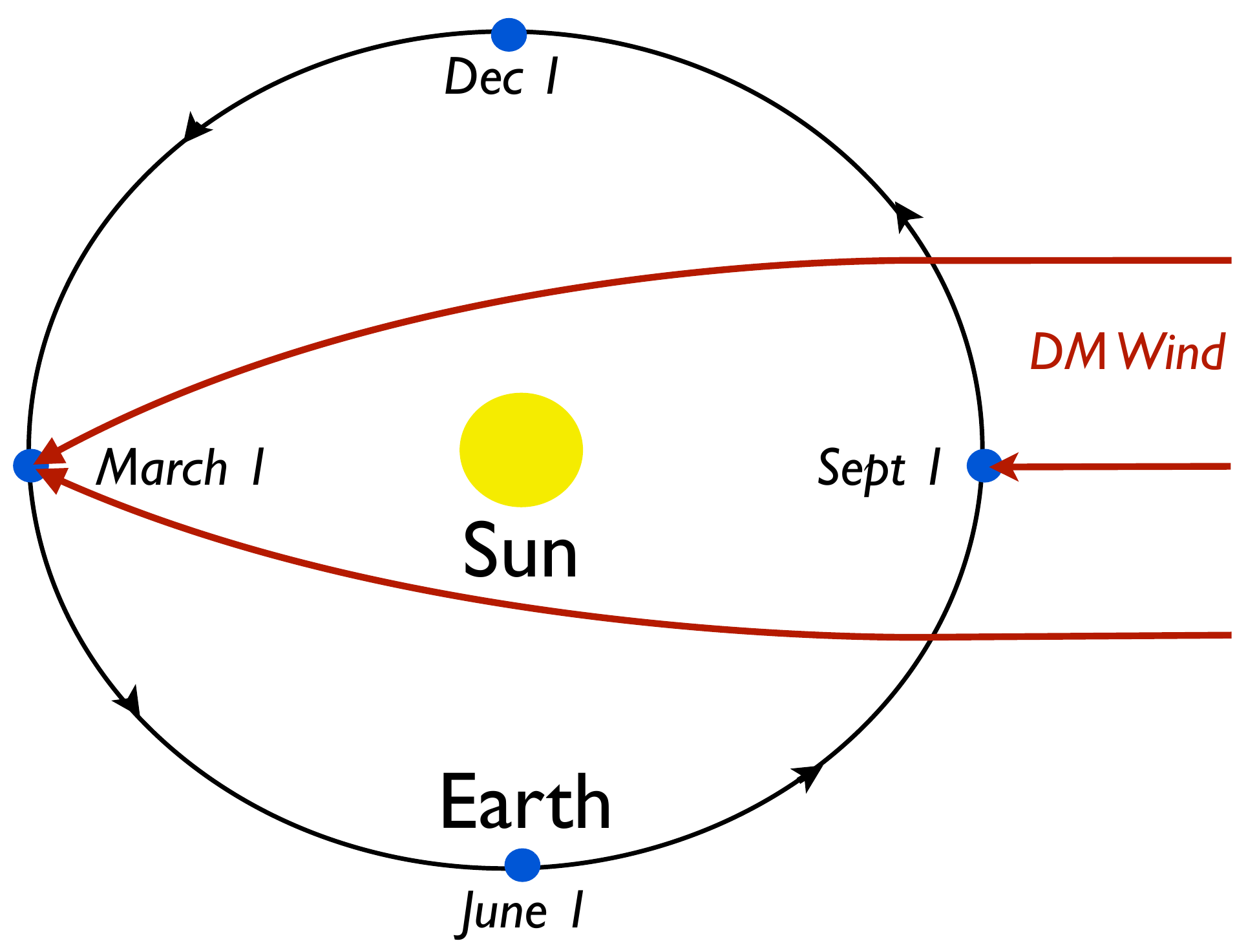} 
   \caption{A schematic representation of the Earth's orbit around the Sun, and the relative orientation of the incident DM wind.  Figure from~\cite{Lee:2013wza}.}
     \label{fig: mod}
\end{figure}

\subsection{Annual Modulation}

Let us return back to (\ref{eq: dRdER}) and discuss the time-dependence of the scattering rate, which arises from boosting the velocity distribution into the lab frame~\cite{Drukier:1986tm}.\footnote{Daily modulation signals can also be relevant.  For example, a daily modulation may be observable in the \emph{direction} of the scattered nucleus, as originally pointed out in~\cite{Spergel:1987kx}.  Another possibility is that the total scattering rate may modulate daily due to the rotation of the Earth about its axis; the exposure needed to detect this is roughly equivalent to that of detecting higher-order modes in the annual modulation.}  The details of the calculations can be found in articles such as~\cite{LewinSmith,Freese:2012xd, Lee:2013xxa}, but we will focus primarily on an illustrative explanation here.  The DM has no preferred direction of motion in the Galactic rest frame.   In the lab frame, however, the DM velocities are oriented opposite to the motion of the Sun (but with roughly equivalent speed).  Therefore, there is a `wind' of DM in the Solar frame, as illustrated in Fig.~\ref{fig: mod}.  In June, the Earth moves towards this wind and an observer sees more high-velocity particles than when the Earth moves away from the wind in December.  As a result, the flux of DM is larger in the summer, compared to the winter,\footnote{The exception is for DM particles with speeds below $\sim$$200$~km/s, which have a phase peaked in winter.  See~\cite{Lee:2013xxa} for further discussion.} resulting in the annual modulation captured by the second term in~(\ref{eq: taylor}).  

The story is not quite this simple, though, and it turns out that an additional effect called `gravitational focusing' may alter the modulation phase~\cite{Lee:2013wza}.  This focusing arises from the fact that a DM particle traveling past the Sun is pulled closer to it by their mutual gravitational attraction.  The net result is an enhancement in the DM phase-space distribution in March, when the Earth is behind the Sun (relative to the wind), as compared to September when it is in front---as illustrated in Fig.~\ref{fig: mod}.  The modulation from gravitational focusing competes with that from the Earth's orbit, and the observed phase is determined by which effect is stronger.  In general, the slower the DM travels, the larger the effect on its trajectory and the more the phase shifts away from June/Dec and closer to Mar/Sep.  Imagine, for instance, that a DM particle is detected and the modulated scattering rate is measured in several energy bins.  The phase in the highest-energy bins will be closely aligned with June/Dec, but will shift closer to Mar/Sep in the lower-energy bins.  The expected energy dependence of the phase is well understood theoretically for DM scattering, and can thus play an important role in distinguishing a potential signal from background.

\section{Application 2:  Indirect Detection} 

Next, we turn to the class of DM searches referred to as indirect detection.  The goal of these experiments is to detect the products of DM annihilation in our Galaxy, or beyond.  While DM annihilation is strongly suppressed after thermal freeze-out, it can still occur today and one can maximize the chance of discovery by searching in regions of very high DM density.  Depending on the theoretical model, the DM can either annihilate directly into a pair of photons, or into other Standard Model states that produce photons in secondary interactions.  The gamma-rays then propagate essentially unperturbed, to (hopefully) be detected by a satellite or ground-based telescope on Earth.  

\subsection{Photon Flux from Annihilations}

We assume that the DM can have multiple annihilation channels, each with velocity-averaged cross section $\langle \sigma_i v \rangle$.  Then, the annihilation rate per particle is 
\be
\sum_i \frac{\rho\left[r(\ell, \psi)\right]}{m_\chi} \times \langle \sigma_i  v \rangle \, ,
\label{eq: perparticle}
\ee
where $r$ is the radial distance between the annihilation event and the Galactic Center---it is a function of the line-of-sight (l.o.s.) distance, $\ell$, which is oriented an angle $\psi$ away from the Galactic plane.  The total annihilation rate in the volume $dV = \ell^2 \, d\ell \, d\Omega$ is obtained by multiplying \eqref{eq: perparticle} by the total number of particles in the volume:
\be
\left( \sum_i \frac{\rho\left[r(\ell, \psi)\right]}{m_\chi} \, \langle \sigma_i  v \rangle \right) \times
\left( \frac{\rho\left[r(\ell, \psi)\right]}{2 \, m_\chi} \, dV \right) \, . 
\label{eq: totalannihilation}
\ee
Note that the factor of two in the denominator comes from the fact that there are two particles involved in every annihilation interaction.  To get the photon flux, we must multiply the annihilation rate of \eqref{eq: totalannihilation} by $dN_i/dE_\gamma$, which describes the number of photons at a given energy $E_\gamma$ produced in the $i^\text{th}$ annihilation channel.  It follows that the differential photon flux $d\Phi/dE_\gamma$ in the observational volume oriented in the direction $\psi$ is 
\be
\frac{d \Phi}{dE_\gamma} \left(E_\gamma, \psi\right) = \frac{1}{4\pi} \, \int_{\Delta\Omega} d\Omega \int_\text{l.o.s.} d \ell\, \rho\left[r(\ell, \psi) \right]^2 \sum_i \frac{\langle \sigma_i v \rangle}{2m_\chi^2} \, \frac{dN_i}{dE_\gamma}  \, ,
\label{eq: flux}
\ee
Importantly, \eqref{eq: flux} must be multiplied by an additional factor of $1/2$ if the DM is not its own antiparticle.

\vspace{0.1in}
\vbox{
\hrule
\vspace{0.1in}
\noindent \textsc{Exercise:} How would (\ref{eq: flux}) change for the case of DM decay rather than annihilation?
\vspace{0.1in} 
\hrule
}
\vspace{0.1in}

All the astrophysical uncertainties in the determination of the flux are absorbed by the $J$-factor, 
\be
J = \frac{1}{\Delta \Omega} \int d\Omega \int_\text{l.o.s.} d \ell\, \rho\left[r(\ell, \psi) \right]^2  \, .\nonumber
\ee
The larger the $J$-factor, the more interesting the astrophysical target is for DM annihilation.  The $J$-factors for dwarf galaxies are roughly $J\sim 10^{19-20}$~GeV$^2$/cm$^5$.  For our nearest neighbor, the Andromeda galaxy, $J\sim10^{20}$~GeV$^2$/cm$^5$.  For our own Galactic Center, $J\sim10^{22-25}$ GeV$^2$/cm$^5$ ($10^{22-24}$) within 0.1$^\circ$(1$^\circ$)~\cite{Profumo:2013yn}.\footnote{It is important to keep in mind that there are potentially large uncertainties on these estimates!}  Choosing an ideal target involves carefully balancing the size of the $J$-factor with the potential backgrounds.  For example, dwarf galaxies are DM-dominated  and therefore some of the cleanest systems to search for DM because they contain very few stars and little gas.  In contrast, a signal from the center of the Galaxy, while enhanced due to the DM density and proximity, has to contend with large systematic uncertainties on the astrophysical backgrounds.  

The particle physics input to the flux is absorbed by the factor of $\frac{\langle \sigma v \rangle}{m_\chi^2} \frac{dN}{dE_\gamma}$.  In many instances, the velocity-averaged cross section can be pulled out of the integral.  However, this cannot be done if the cross section depends strongly on velocity, as is the case for $p$-wave annihilation or Sommerfeld enhancements (which we will come back to later).  The kinematics of the annihilation event determine the basic properties of the photon energy spectrum.  Consider, first, the case where the DM annihilates directly into one or two photons: $\chi \chi \rightarrow \gamma X$, where $X = \gamma, Z, H$ or some other neutral state.  In the non-relativistic limit, energy conservation gives
\be
2 \, m_\chi = E_\gamma + \sqrt{E_\gamma^2 + m_X^2} \longrightarrow E_\gamma \approx m_\chi \left( 1- \frac{m_X^2}{4 m_\chi^2} \right) \, , \nonumber
\ee
where $E_\gamma$ is the energy of the outgoing photon in the center-of-mass frame and $m_X$ is the mass of the $X$ state.  To get the expression on the right-hand side, we assume that the energy of the outgoing photon is  $k = m_\chi + \delta$ and expand in the mass difference $\delta$.  The $\gamma \gamma$ final state results in a monochromatic energy line at the DM mass.  For a $\gamma Z$ final state, the gamma line is still monochromatic, but is shifted to lower energies.  

\begin{figure}[tb] 
   \centering
   \includegraphics[width=4in]{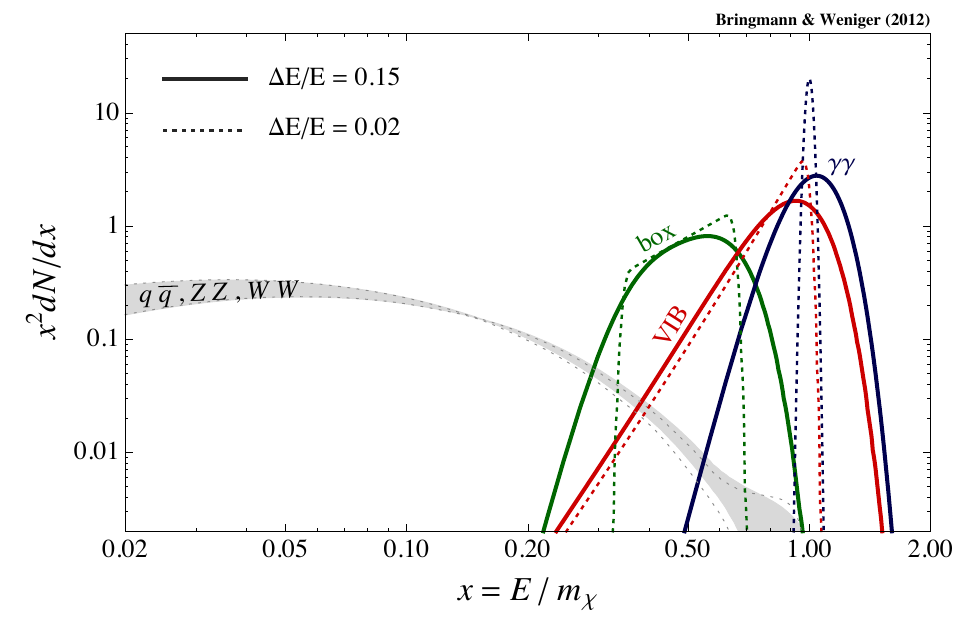} 
   \caption{Illustration of the photon energy spectrum for the $\gamma\gamma$ final state without (blue) and with (red) virtual internal bremsstrahlung.  The box spectrum (green) can be produced if the DM annihilates to a new state, that then decays to photons, as described in the text.  The dotted versus solid lines compare two separate energy resolutions: $\Delta E/E=$ 0.02 and 0.15, respectively.  The spectrum for photons resulting from the annihilation into gauge bosons and quarks is shown by the gray band.  Figure from~\cite{Bringmann:2012ez}.}
   \label{fig: gamma_spectrum}
\end{figure}

The blue lines in Figure~\ref{fig: gamma_spectrum} show the energy spectrum for a $\gamma\gamma$ final state where the measured energy resolution is $\Delta E/E = 0.15$ (solid) or $0.02$ (dotted).  The observation of such a gamma-ray `line' would be spectacular evidence for DM annihilation.  However, the production of a pair of gamma-rays is typically loop-suppressed (and therefore sub-dominant) in many theories.   The red lines in Fig.~\ref{fig: gamma_spectrum} illustrate how the spectrum changes if photons are radiated off of virtual charged particles in the loop.  Such `virtual internal bremsstrahlung' results in a broadening of the line towards lower masses, though the spectrum still cuts off at the DM mass.  The green lines in Fig.~\ref{fig: gamma_spectrum} illustrate the box spectrum, which arises when the DM annihilates to a new state $\phi$ (\emph{e.g.}, $\chi\chi \rightarrow \phi \phi$) that then decays to a photon pair ($\phi \rightarrow \gamma \gamma$)~\cite{Ibarra:2012dw}.

Another possibility is that the DM annihilates to leptons, gauge bosons, or quarks, which may produce secondary photons either through final-state radiation or in the shower of their decay products.  The photon energy spectrum $dN/dE_\gamma$ depends on the exact details of the final-state radiation, and must be determined with Monte Carlo tools like \texttt{Pythia8}~\cite{Sjostrand:2007gs}.\footnote{For recipes on calculating DM annihilation signals, see the \emph{Poor Particle Physicist Cookbook for Dark Matter Indirect Detection}~\cite{PPPC}.}  In the case of secondary photon production, the energy spectrum does not have a very distinctive shape, and one must search for a continuum excess over the background.  The gray band in Fig.~\ref{fig: gamma_spectrum} shows an example of the spectrum for annihilation to quarks or gauge bosons.
\begin{figure}[tb] 
   \centering
   \includegraphics[width=6.5in]{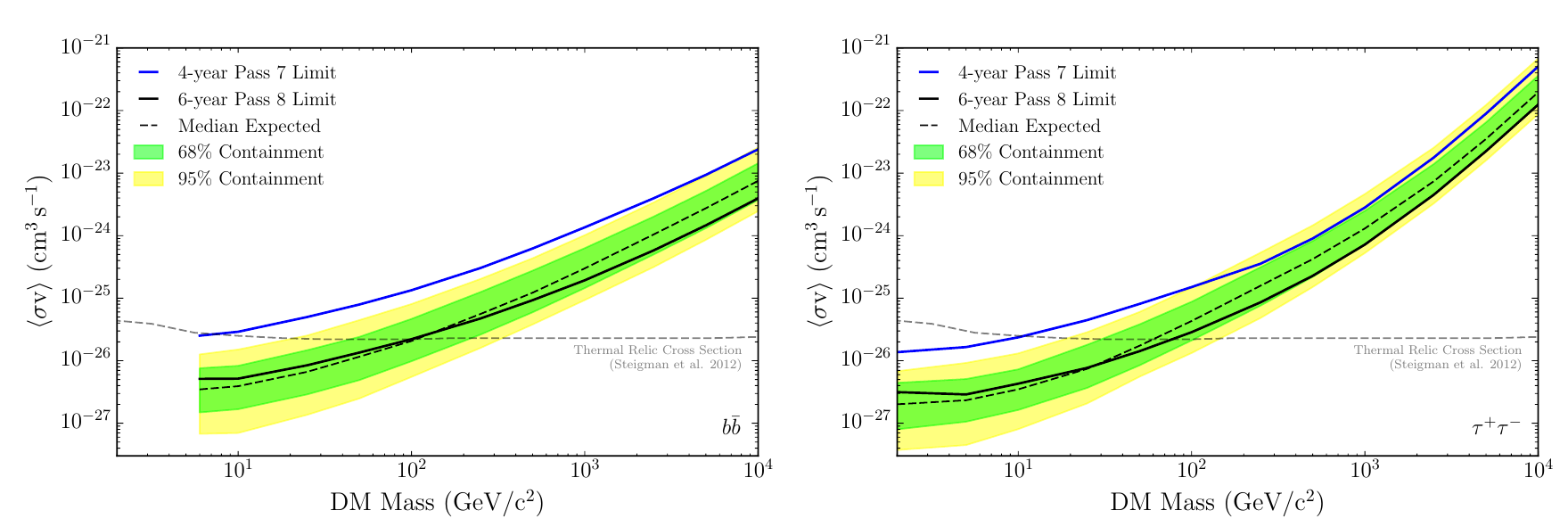} 
   \caption{\emph{Fermi} LAT limits on DM annihilation into $b\bar{b}$ (left) and $\tau^+ \tau^-$ (right) final states.  The dashed black line is the expected bound with 68\% and 95\% contours shown in green and yellow, respectively.  The solid black line is the observation with six-year Pass 8 data.  Figure from~\cite{Ackermann:2015zua}.}
   \label{fig: dwarf}
\end{figure}

The details of the annihilation mechanism are buried in the velocity-averaged cross section $\langle \sigma v \rangle$.  This cross section is the same in many simple models as what appears in the relic density calculation.  As a result, one can argue that indirect detection is the best probe of the thermal DM hypothesis.  In addition, we automatically have an interesting target scale for the cross section: $3\times10^{-26}$~cm$^3$\,s$^{-1}$.  This regime is currently being probed by the best gamma-ray observatories today.  For example, the \emph{Fermi} Large Area Telescope has searched for signals of DM annihilation in the Milky Way's dwarf galaxies~\cite{Ackermann:2015zua}.  Figure~\ref{fig: dwarf} shows the limits from their most recent analysis, assuming annihilation to $b\bar{b}$ (left) and $\tau \bar{\tau}$ (right). (Remember, the assumption of the final decay products affects $dN/dE_\gamma$.)  Such limits are typically presented in terms of the velocity-averaged cross section and the DM mass.  The dashed black line shows the median expected limit with 68\% and 95\% confidence bands in green and yellow, respectively.  The observed limit with six years of data is shown by the solid black line.  The horizontal dotted black line shows the thermal relic cross section, to guide the eye.  Notice that the observed bounds are starting to push into the parameter regime that is highly motivated for WIMP dark matter.   

\subsection{Sommerfeld Enhancement}

\begin{figure}[tb] 
   \centering
   \includegraphics[width=6.5in]{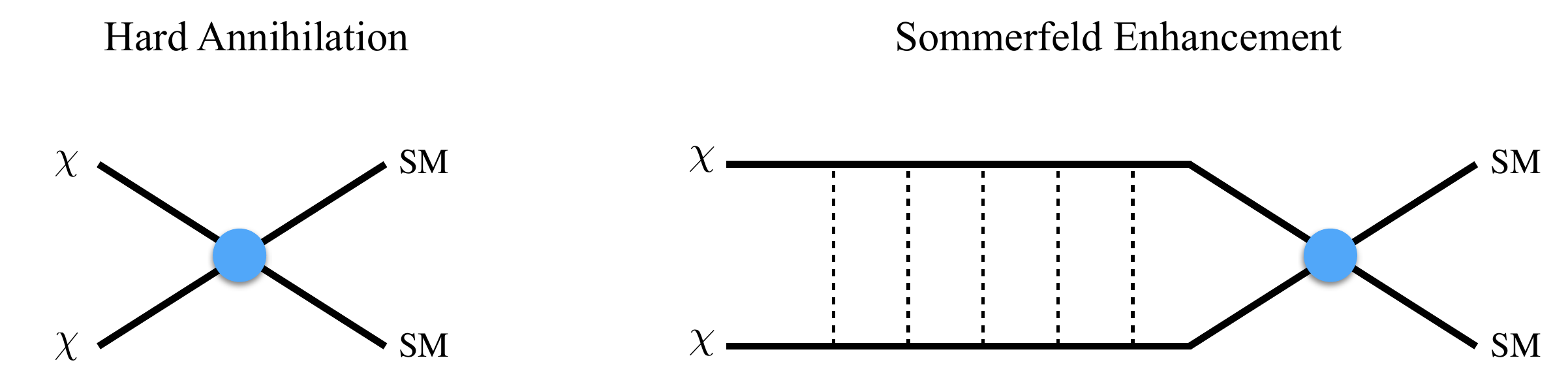} 
   \caption{A schematic illustration of the annihilation of two DM particles, $\chi$, into Standard Model (SM) final states without (left) and with (right) Sommerfeld enhancement.  }
   \label{fig: sommerfelddiagram}
\end{figure}

As an example of how this story can change if $\langle \sigma v \rangle$ is no longer constant in velocity, let us consider scenarios where DM self-interactions are allowed.  In such cases, some very interesting non-relativistic effects can arise that drastically alter the energy spectrum for the annihilation process.  In certain instances, this can make the difference between discovering the DM or not~\cite{Cirelli:2007xd, ArkaniHamed:2008qn}.  The general idea is illustrated in Fig.~\ref{fig: sommerfelddiagram}.  Assume that the annihilation of the DM into Standard Model final states is a localized interaction---say, at the origin.  If there are no self-interactions between the DM particles, then the annihilation process looks like the left panel of Fig.~\ref{fig: sommerfelddiagram}.  In this case, the probability of finding the DM particles at the origin is just $\left|\psi_0(0)\right|^2$, where $\psi_0$ is the incoming wave function and a solution to the non-relativistic Schr\"{o}dinger equation.   

If self-interactions are allowed by the theory, then one possibility is that the DM particle can interact with itself via a long-range force before annihilating.  For example, if we introduce a new scalar $\phi$ that couples to the DM via $\bar{\chi} \chi \phi$, then the two $\chi$ legs of the diagram can exchange multiple $\phi$ states before the hard annihilation occurs at the origin.  The exchange of multiple mediators alters the wave function of the incoming DM particles so that the probability of finding them at the annihilation site is now $| \psi(0) |^2$, where $\psi$ is the modified wave function in the presence of the interaction potential.  This is known as the Sommerfeld enhancement. 

The Sommerfeld enhancement is defined as the ratio of probabilities of finding the DM at the origin in the presence of the potential, relative to no potential:
\be
S = \frac{|\psi(0)|^2}{|\psi_0(0)|^2} \, . \nonumber
\ee
To calculate $\psi(r)$, one must evaluate the ladder diagram in the right panel of Fig.~\ref{fig: sommerfelddiagram}.  This diagram is non-perturbative and determining $\psi(0)$ would be a much more challenging problem if we could not treat the DM system non-relativistically.  Fortunately, we can and $\psi(r)$ is obtained by solving the Schr\"{o}dinger equation for the non-relativistic effective potential that describes the interaction~\cite{Hisano:2003ec, Hisano:2004ds, Cassel:2009wt, Iengo:2009ni}.

Let us jump into the details now and explore the interesting phenomenology of such processes.  The wave function, $\psi$, for two-particle scattering is described by the time-independent Schr\"{o}dinger equation 
\be
-\frac{1}{2 m_\chi} \nabla_1^2 \,\psi - \frac{1}{2m_\chi} \nabla_2^2 \,\psi + V(\mathbf{r}_1, \mathbf{r}_2) \, \psi = E_\text{lab} \, \psi \, , \nonumber
\ee
where the two DM particles each have mass $m_\chi$ and $E_\text{lab}$ is the energy in the lab frame.  In the center-of-mass frame, this becomes
\be
-\frac{1}{2 \mu}  \nabla_r^2 \, \psi + V(r) \, \psi = E \, \psi \, ,
\label{eq: Schrod}
\ee
where $\mu =  m_\chi/2$ is the reduced mass, $\mathbf{r} = \mathbf{r}_1 - \mathbf{r}_2$ is the separation of the two particles, and $E= \frac{1}{2} \mu v^2$ is the center-of-mass energy.  Note the change in notation---$\psi$ now refers to the wave function for a single particle of mass $\mu$ (that describes the full non-relativistic two-DM state) scattering off the potential $V(r)$.   We expand the wave function in terms of partial waves:
\be
\psi(r, \theta, \phi) = \sum_l \tilde{A}_l \, R_{kl}(r) \, Y_l^m(\theta, \phi) =  \sum_l A_l \, R_{kl}(r) \, P_l \left( \cos\theta\right) .  
\label{eq: wavefunction}
\ee
The solutions are separable and the angular equation gives $Y_l^m (\theta, \phi)$ proportional to the associated Legendre function.  Note that the azimuthal dependence vanishes due to the symmetry of the problem.  We are left with the following equation for the radial term, $R_{kl}(r)$:
\be
\frac{d}{dr}\left( r^2 \, \frac{d R_{kl}}{dr} \right) - m_\chi r^2 \Big\{V(r) - 
E \Big\} R_{kl} = l(l+1) R_{kl} \, \nonumber
\ee
To simplify this further, we apply the change of variables $u_{kl}(r) \equiv r R_{kl}(r)$, and get
\be
\frac{d^2 u_{kl}}{dr^2} - m_\chi \left\{ V(r) - E + \frac{1}{m_\chi} \frac{l(l+1)}{r^2} \right\} u_{kl}= 0 \, .
\label{eq: radial}
\ee
For concreteness, let us focus on the Yukawa potential,
\be
V(r) = \frac{\alpha}{r}e^{-m_\phi r} \, , \nonumber
\ee
which arises when the interaction is mediated by a boson of mass $m_\phi$.  Notice that the potential $V(r)$ can be ignored in (\ref{eq: radial}) as $r\rightarrow0$ so long as it blows up less rapidly than $1/r^2$.  In this limit,
\be
\frac{d^2 u_{kl}}{dr^2}  = \left[ \frac{l(l+1)}{r^2} - k^2 \right] u_{kl} \,  \longrightarrow R_{kl}(r) \propto j_{l}(kr) \, , \nonumber
\label{eq: ul}
\ee
where $k \equiv \sqrt{m_\chi E}$ and $j_l(x)$ is the spherical Bessel function of order $l$.  When $x\rightarrow 0$, $j_l(x) \sim x^l$; therefore, only the $l = 0$ term is relevant at the origin.  As a result, we can focus on the scenario of $s$-wave scattering and consider only the $l=0$ partial wave.  To simplify notation, we take $u_{k, l=0} = u_k$ in the following.

\begin{figure}[tb] 
   \centering
   \includegraphics[width=4.5in]{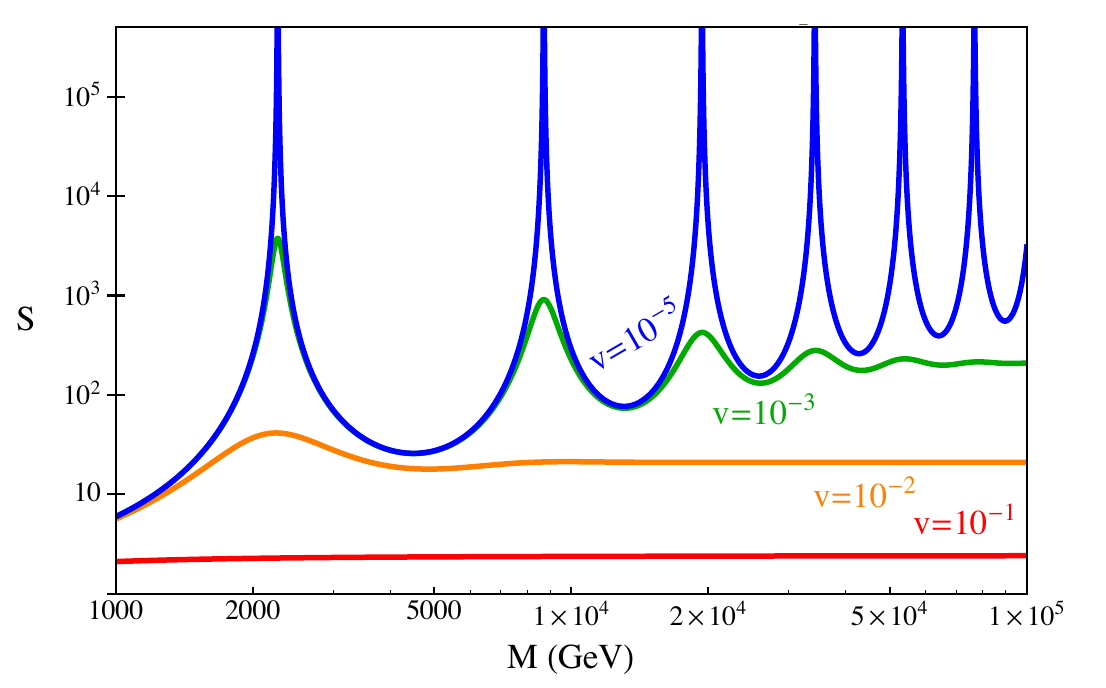} 
   \caption{The Sommerfeld enhancement for the attractive ($\alpha<0$) Yukawa potential with $m_\phi =  90$~GeV and $\alpha = 1/30$.  For a given DM mass $m_\chi = M$, as the relative DM velocity decreases, the enhancement increases.  Resonance peaks appear at very small velocities when the DM forms bound states.
  Figure from~\cite{Bellazzini:2013foa}.}
   \label{fig: sommerfeld}
\end{figure}

The radial Schr\"{o}dinger equation for arbitrary $r$ becomes
\be
\frac{d^2 u_k}{dr^2} - m_\chi \, V(r) \, u_k = - k^2 \, u_k \, ,
\label{eq: radial_l0}
\ee
which can be solved numerically for $u_k(r)$ subject to the boundary condition that the outgoing wave be spherical.  Figure~\ref{fig: sommerfeld} shows the result of such an evaluation.  There are a few pertinent features of the Sommerfeld enhancement that are worth highlighting.  First, the enhancement is velocity-dependent and increases as the relative velocity decreases from $10^{-1}$ to $10^{-3}$.  At even smaller relative velocities, a distinct resonance structure appears; for masses that yield such resonances, the Sommerfeld factor is strongly enhanced.  It turns out to be fairly straightforward to understand this behavior if we study (\ref{eq: radial_l0}) in certain limits where analytic solutions are possible.

In the limit where $m_\phi \rightarrow 0$, we obtain the Coulomb potential and (\ref{eq: Schrod}) can be solved exactly by transforming to parabolic coordinates:
\be
\xi = r-z \quad \text{ , } \quad \eta = r + z \quad \text{ and } \quad \phi = \tan^{-1}\left(\frac{y}{x}\right)  \, .\nonumber
\ee    
The Schr\"{o}dinger equation becomes
\be
\left[ -\frac{1}{2\mu} \frac{4}{\xi + \eta} \left( \frac{\partial}{\partial \xi} \xi \frac{\partial}{\partial \xi}+ \frac{\partial}{\partial \eta} \eta \frac{\partial}{\partial \eta}  + \frac{\xi + \eta}{4\xi \eta} \frac{\partial^2}{\partial \phi^2} \right) - \frac{2 \, \alpha}{\xi + \eta} \right] \psi = \frac{k^2}{2 \mu} \psi \, . \nonumber
\ee
Because the scattering potential is azimuthally symmetric, we need only consider solutions of the form
\be
\psi(\xi, \eta) = e^{\frac{i}{2} k(\eta - \xi)} \, \Phi(\xi) = e^{i k z} \, \Phi(r-z) \, , \nonumber
\ee
so that  
\be
\left[ \xi \frac{\partial^2}{\partial \xi^2} + (1-ik\xi) \frac{\partial}{\partial \xi} + \alpha \, \mu \right] \Phi = 0  \, . \nonumber
\ee
The solutions to equations of this form are known as the confluent hypergeometric functions:
\be
\Phi(\xi) = A \,\, M(i\lambda;\,1;\, ik \xi) \longrightarrow \psi(r,z)  = A \,\,M (i\lambda;\,1;\, ik (r-z) )\, e^{ikz} \, , \nonumber 
\ee
where $\lambda = \alpha \mu/ k $.  The hypergeometric function has the property that $M(a;\, b;\, 0) = 1$, so the probability of finding the particle at the origin is $\left| \psi \right|^2 =   |A|^2$.  The probability of finding the particle in the incident beam is 
\be
\left| \psi (r, z) \right|^2 \rightarrow \frac{ \left|A\right|^2 e^{\pi \lambda}}{\left|\Gamma (1- i \lambda) \right|^2} \, , \nonumber 
\ee
where we used the asymptotic form of the hypergeometric function ($|k \, \xi |\rightarrow \infty$).  Note that this is equivalent to the probability of finding the particle at the origin if there were no potential.  Because 
\be
\Gamma(1+iy) \, \Gamma(1-iy) = \left| \Gamma(1 + i y) \right|^2 = \frac{2 \, \pi \, |y| \, e^{\pm \pi |y|}}{\pm (e^{\pm 2 \pi |y|} -1 )} \, ,\nonumber
\ee
it follows that
\be
S = \left| \frac{\psi(0)} {\psi_0(0)}\right|^2 = \frac{2 \pi | \lambda |}{\pm e^{\pm 2\pi |\lambda|}-1} = \left| \frac{\alpha}{v} \right| \frac{2 \pi}{\pm \left( e^{\pm 2 \pi |\alpha|/v} -1 \right)} \, . \nonumber
\ee
For both repulsive and attractive interactions, $S\rightarrow 1$ as $v\rightarrow  \infty$, which makes intuitive sense as the two particles pass by each other too quickly to be affected by the potential.  In the repulsive case ($\alpha > 0$), $S\rightarrow 0$ as $v\rightarrow 0$ because the potential pushes the two particles apart, inhibiting their interaction.  However, in the attractive case ($\alpha < 0$), $S\rightarrow 2\pi \alpha/v$ and the enhancement grows as the relative velocity decreases!  

This enhancement does not grow arbitrarily large with decreasing $v$, however.  This is because the kinetic energy term in (\ref{eq: radial_l0}) eventually becomes subdominant to the potential attraction between the two particles and bound states form.  To better quantify when this transition occurs, we rewrite the Schr\"{o}dinger equation by introducing the new variable $x\equiv m_\phi r$ and expanding the Yukawa potential for $x\ll1$:
\be
\frac{d^2 u_k}{dx^2} + \frac{\alpha }{x} \frac{m_\chi}{m_\phi} \, u_k = \left(- \frac{k^2}{m_\phi^2} + \alpha \, \frac{m_\chi}{m_\phi} \right) u_k \, ,
\label{eq: urewritten} 
\ee
where we assume an attractive ($\alpha < 0$) potential since that is the only case where the resonances arise.
The Coulomb approximation holds so long as $k^2 \gg \alpha \, m_\chi \, m_\phi \rightarrow v \gg 2 \, \sqrt{\alpha m_\phi/m_\chi}$, as then the second term on the right-hand side of (\ref{eq: urewritten}) can be ignored.  For smaller relative velocities,
\be
\frac{d^2 u_k}{dx^2} + \frac{\alpha }{x} \frac{m_\chi}{m_\phi} \, u_k = \alpha \, \frac{m_\chi}{m_\phi} \, u_k \, ,
\label{eq: boundstate}
\ee
and the potential attraction is so strong that the particles form bound states.  The form of (\ref{eq: boundstate}) is reminiscent of the radial equation for the Hydrogen atom.  By analogy,
\be
\alpha \frac{m_\chi}{m_\phi} \sim n^2 \quad \longrightarrow  \quad m_\chi \sim \frac{m_\phi}{\alpha} \, n^2 \, \quad \text{for } n=1,2,3\ldots \nonumber
\ee
Therefore, the DM bound states form only for discrete values of the DM masses in the ratio $1:4:9$ and so on~\cite{Lattanzi:2008qa}.  The resonance peaks in Fig.~\ref{fig: sommerfeld} do indeed satisfy this relation.  

Having built intuition for the Sommerfeld mechanism in the case of the Yukawa potential, let us now briefly discuss the procedure that one must follow to evaluate $S$ for arbitrary potentials.  This calculation is a standard non-relativistic scattering problem and is reviewed in many quantum mechanics texts, so we will only outline the procedure here.  In general, the solution to the Schr\"{o}dinger equation (\ref{eq: wavefunction}) must take the asymptotic form
\be
\psi(r,\theta) \rightarrow e^{i k z} + f(\theta) \frac{e^{i k r}}{r} , 
\label{eq: asymptotic}
\ee
assuming that the incoming wave is described by $\psi_0(r) = e^{i k z}$ and the scattered wave is spherical.  
In the asymptotic limit, the radial wave function $R_{kl}(r)$ is
\be
R_{kl}(r) \sim \frac{2}{r} \sin \left( kr - \frac{\pi}{2} l + \delta_l \right) \, , \nonumber 
\ee 
where $\delta_l$ accounts for potential phase shifts from the scattering potential.\footnote{Note that writing down the asymptotic form of $R_{kl}(r)$ requires that $V(r) \rightarrow 0$ faster than $1/r$ as $r\rightarrow\infty$.  This assumption is not valid for the Coulomb potential, which is why we needed to solve the Schr\"{o}dinger equation exactly in that case.}  It can be shown that
\be
A_l = \frac{1}{2k} i^l (2l+1) e^{i\delta_l} \nonumber
\ee
in order for (\ref{eq: wavefunction}) to take the form (\ref{eq: asymptotic}) in the asymptotic limit.  Therefore, the wave function solution is 
\be
\psi(r, \theta) =  \frac{1}{k} \sum_l  i^l (2l + 1) e^{i\delta_l} \, P_l \left( \cos\theta\right) \, R_{kl}(r)  \, .  \nonumber
\ee
Remembering that only the $l=0$ term contributes at the origin, the expression for the Sommerfeld enhancement becomes
\be
S = \left| \frac{R_{k,l=0}(0)}{k} \right|^2 \nonumber 
\ee
and the problem reduces to evaluating the $l=0$ partial wave of the radial wave function at the origin.

Our discussion so far has only dealt with the case of a one-state system, where each `rung' on the ladder diagram is the same: $\chi \bar{\chi} \rightarrow \chi \bar{\chi}$.  In some models, however, new states that are nearly degenerate to the DM may exist, in which case they could also contribute to the interactions~\cite{Slatyer:2009vg}.  For instance, consider what happens when a new charged particle $\chi^\pm$ that is nearly degenerate with the DM is introduced.  In this case, the ladder diagram can be built up from different rungs:
\be
\chi \bar{\chi}  \rightarrow \chi  \bar{\chi} \, , \quad \chi \bar{\chi} \rightarrow \chi^+ \chi^- \, , \quad  \chi^+ \chi^- \rightarrow \chi \bar{\chi}  \, \quad \text{ and } \quad  \chi^+ \chi^- \rightarrow \chi^+ \chi^- \, ,\nonumber
\ee
where the mediator in each rung varies depending on the interacting particles.  For this two-state system, the Schr\"{o}dinger equation is the same as above, except that the potential $\mathbf{V}(r)$ and radial wave function $\mathbf{R}_{kl}$ become $2\times2$ matrices of the form
\be
\mathbf{V}(r) = \left(\begin{array}{c c}
V_{00}^{00}(r) & V_{00}^{+-}(r)  \\
V_{+-}^{00}(r) & V_{+-}^{-+}(r) \end{array}\right)   \quad \text{ and } \quad 
\mathbf{R}_{kl}(r) = \left(\begin{array}{c c}
R_{00}^{00}(r) & R_{00}^{+-}(r)  \\
R_{+-}^{00}(r) & R_{+-}^{-+}(r) \end{array}\right) \nonumber \, ,
\ee
where each $V_{ij}^{kl}(r)$ describes the effective potential between the $ij$ initial state and $kl$ final state, given by the radial wave function $R_{ij}^{kl}(r)$.  Note that `0' is shorthand for $\chi$ and `$\pm$' is shorthand for $\chi^\pm$.  The wino is a supersymmetric DM candidate that provides a classic example of a two-state system where Sommerfeld enhancement plays an integral role~\cite{Cohen:2013ama, Fan:2013faa}.\footnote{See~\cite{Bauer:2014ula, Baumgart:2014vma, Ovanesyan:2014fwa} for a discussion of important corrections to the Sommerfeld calculations that arise in this example.}  In order for the wino to be a thermal relic, its mass must be $\sim$3~TeV.  The possibility of observing the final-state annihilation products for such heavy winos is very challenging.  However, the cross section is Sommerfeld enhanced due to the exchange of $W^\pm$ and $Z^0$ bosons.  Indeed, it is only because of this enhancement that Cherenkov telescope arrays have the sensitivity to exclude certain regions of the parameter space.  

\section{Summary}

The goal of these lectures was to provide the reader with the basics of DM theory.  As we have seen, the last few decades have brought great progress in the understanding of DM.  Well-motivated hypotheses, such as WIMPs, have provided a starting point for experimental exploration and current experiments are reaching the necessary sensitivities to discover or exclude these candidates.  However, weak-scale DM is not a guarantee; as discussed, a broad range of interactions and mass scales are allowed.  A diverse experimental program is therefore crucial for success.  These lectures focused specifically on direct and indirect detection experiments as examples, but there are many more---including collider, axion, and beam-dump experiments.  Each approach is complementary and has the potential to provide a unique window into the dark sector.   Hopefully, the interested reader will feel emboldened to pursue these topics in greater depth, ready for whatever surprises lay ahead.

\section{Acknowledgements}
I would like to thank the TASI organizers for the invitation to lecture at the 2015 summer program; it was a great pleasure to interact with the enthusiastic and motivated students at the school.  I am also grateful to T.~Cohen, R.~D'Agnolo, Y.~Kahn, and S. Mishra Sharma for carefully reviewing the manuscript and providing helpful feedback.  This work is supported by the DoE under grant Contract Number DE-SC0007968.

\bibliographystyle{ssg}
\bibliography{ws-rv-sample}

\end{document}